\documentclass[longauth]{aa}  

\usepackage{natbib}
\bibpunct{(}{)}{;}{a}{}{,} 
\usepackage[colorlinks=true,linkcolor=blue,citecolor=blue]{hyperref} 

\usepackage{placeins} 
\usepackage{multicol} 

\usepackage{academicons}
\usepackage{xcolor}
\newcommand{\orcid}[1]{\href{https://orcid.org/#1}{\textcolor[HTML]{A6CE39}{\aiOrcid}}}

\newcommand{\toidos}{TOI-2081 }
\newcommand{\toidosb}{TOI-2081b }

\newcommand{\toicuatro}{TOI-4479 }
\newcommand{\toicuatrob}{TOI-4479b }

\usepackage{graphicx}
\usepackage{txfonts}
\begin{document}

   \title{A hot sub-Neptune in the desert and a temperate super-Earth around faint M dwarfs}

   \subtitle{Color validation of \toicuatrob and \toidosb}

   \author{E.~Esparza-Borges\inst{\ref{IAC}}\fnmsep\inst{\ref{ull}} \thanks{emma.esparza.borges@iac.es}
          \and
          H.~Parviainen\inst{\ref{IAC}}\fnmsep\inst{\ref{ull}}
          \and
          F.~Murgas\inst{\ref{IAC}}\fnmsep\inst{\ref{ull}}
          \and
          E.~Pall\'e\inst{\ref{IAC}}\fnmsep\inst{\ref{ull}}
          \and
          A.~Maas\inst{\ref{uheidelberg}}\fnmsep\inst{\ref{konigstuhl}}
          \and
          G.~Morello\inst{\ref{IAC}}\fnmsep\inst{\ref{ull}}
          \and
          M.R.~Zapatero-Osorio\inst{\ref{cab}}
          \and
          K.~Barkaoui\inst{\ref{astrob_belgium}}\fnmsep\inst{\ref{MIT_EAPS}}\fnmsep\inst{\ref{IAC}}
          \and
          N.~Narita\inst{\ref{KomabaInst}}\fnmsep\inst{\ref{AstrobMitaka}}\fnmsep\inst{\ref{IAC}}
          \and
          A.~Fukui\inst{\ref{KomabaInst}}\fnmsep\inst{\ref{IAC}}
          \and
          N.~Casasayas-Barris\inst{\ref{Leiden}}
          \and
          M.~Oshagh\inst{\ref{IAC}}\fnmsep\inst{\ref{ull}}
          \and
          N.~Crouzet\inst{\ref{Leiden}}
          \and
          D.~Gal\'an\inst{\ref{ull}}
          \and
          G.E.~Fern\'andez\inst{\ref{IAC}}
          \and
          T.~Kagetani\inst{\ref{Dep_multi_tokio}}
          \and
          K.~Kawauchi\inst{\ref{IAC}}\fnmsep\inst{\ref{ull}}
          \and
          T.~Kodama\inst{\ref{KomabaInst}}
          \and
          J.~Korth\inst{\ref{DepSpace_sweeden}}
          \and
          N.~Kusakabe\inst{\ref{AstrobMitaka}}\fnmsep\inst{\ref{NAOJ}}
          \and
          A.~Laza-Ramos\inst{\ref{IAC}}
          \and
          R.~Luque\inst{\ref{IAA-csic}}
          \and
          J.~Livingston\inst{\ref{AstrobMitaka}}\fnmsep\inst{\ref{NAOJ}}\fnmsep\inst{\ref{DepaAstro_Tokio}}
          \and
          A.~Madrigal-Aguado\inst{\ref{IAC}}
          \and
          M.~Mori\inst{\ref{DepaAstro_Tokio}}
          \and
          J.~Orell-Miquel\inst{\ref{IAC}}\fnmsep\inst{\ref{ull}}
          \and
          M.~Puig-Subir\`a\inst{\ref{IAC}}
          \and
          M.~Stangret\inst{\ref{IAC}}\fnmsep\inst{\ref{ull}}
          \and
          Y.~Terada\inst{\ref{Taiwan_InstituteAA}}\fnmsep\inst{\ref{Taiwan_DepartmentA}}
          \and
          N.~Watanabe\inst{\ref{Dep_multi_tokio}}
          \and
          Y.~Zou\inst{\ref{Dep_multi_tokio}}
          \and
          A.~Baliga Savel\inst{\ref{uMaryland}}
          \and
          A.A.~Belinski\inst{\ref{sternberg}}
          \and
          K.~Collins\inst{\ref{CfA}}
          \and
          C.D.~Dressing\inst{\ref{Berkeley}}
          \and
          S.~Giacalone\inst{\ref{Berkeley}}
          \and
          H.~Gill\inst{\ref{Berkeley}}
          \and
          M.V.~Goliguzova\inst{\ref{sternberg}}
          \and
          M.~Ikoma\inst{\ref{Div_science_mitaka}}
          \and
          J.M.~Jenkins\inst{\ref{NASA_Ames}}
          \and
          M.~Tamura\inst{\ref{AstrobMitaka}}\fnmsep\inst{\ref{DepaAstro_Tokio}}\fnmsep\inst{\ref{NAOJ}}
          \and
          J.D.~Twicken\inst{\ref{SETI}}
          \and
          G.R.~Ricker\inst{\ref{KavliMIT}}\fnmsep\inst{\ref{PhysicsMIT}}
          \and
          R.P.~Schwarz\inst{\ref{VoorheesvilleObs}}
          \and
          S.~Seager\inst{\ref{KavliMIT}}\fnmsep\inst{\ref{PhysicsMIT}}\fnmsep\inst{\ref{MIT_Aeronautics}}\fnmsep\inst{\ref{MIT_EAPS}}
          \and
          A.~Shporer\inst{\ref{KavliMIT}}
          \and
          R.~Vanderspek\inst{\ref{KavliMIT}}\fnmsep\inst{\ref{PhysicsMIT}}
          \and
          J.~Winn\inst{\ref{Princeton}}
          }

   \institute{Instituto de Astrof\'isica de Canarias (IAC), E-38200 La Laguna, Tenerife, Spain\label{IAC}
         \and
         Departamento de Astrof\'isica, Universidad de La Laguna (ULL), E-38206 La Laguna, Tenerife, Spain\label{ull}
         \and
         Department for Physics and Astronomy, University of Heidelberg, Germany\label{uheidelberg}
         \and
         Landesternwarte Königstuhl (LSW), Zentrum für Astronomie der Universität Heidelberg, Königstuhl 12, D-69117 Heidelberg, Germany\label{konigstuhl}
         \and
         Centro de Astrobiologia (CSIC-INTA), Carretera de Ajalvir km 4, 28850 Torrejon de Ardoz, Madrid, Spain\label{cab}
         \and
         Leiden Observatory, Leiden University, Postbus 9513, 2300 RA, Leiden, The Netherlands\label{Leiden}
         \and
         Center for Astrophysics \textbar \ Harvard \& Smithsonian, 60 Garden Street, Cambridge, MA 02138, USA\label{CfA}
         \and
         Department of Astronomy, University of California Berkeley, Berkeley, CA 94720, USA\label{Berkeley}
         \and
         Department of Astronomy, University of Maryland, College Park, College Park, MD 20742 USA\label{uMaryland}
         \and
         Department of Space, Earth and Environment, Astronomy and Plasma Physics, Chalmers University of Technology, 412 96 Gothenburg, Sweden\label{DepSpace_sweeden}
         \and
         Instituto de Astrof\'isica de Andaluc\'ia (IAA-CSIC), Glorieta de la Astronom\'ia s/n, 18008 Granada, Spain\label{IAA-csic}
         \and
         Sternberg Astronomical Institute, Lomonosov Moscow State University, 13 Universitetski prospekt, 119992 Moscow, Russia. University\label{sternberg}
         \and
         Astrobiology Research Unit, Université de Liège, 19C Allée du 6 Août, 4000 Liège, Belgium\label{astrob_belgium}
         \and
         Patashnick Voorheesville Observatory, Voorheesville, NY 12186, USA\label{VoorheesvilleObs}
         \and
         Department of Physics, Massachusetts Institute of Technology, Cambridge, MA 02139, USA\label{KavliMIT}
         \and
         MIT Kavli Institute for Astrophysics and Space Research, Cambridge, MA 02139, USA \label{PhysicsMIT}
         \and
         Komaba Institute for Science, The University of Tokyo, 3-8-1 Komaba, Meguro, Tokyo 153-8902, Japan\label{KomabaInst}
         \and
         Astrobiology Center, 2-21-1 Osawa, Mitaka, Tokyo 181-8588, Japan\label{AstrobMitaka}
         \and
         Department of Astronomy, The University of Tokyo, 7-3-1, Hongo, Bunkyo-ku, Tokyo 113-0033, Japan\label{DepaAstro_Tokio}
         \and
         National Astronomical Observatory of Japan, 2-21-1 Osawa,Mitaka, Tokyo 181-8588, Japan\label{NAOJ}
         \and
         Division of Science, National Astronomical Observatory of Japan, 2-21-1 Osawa, Mitaka, Tokyo 181-8588, Japan\label{Div_science_mitaka}
         \and
         SETI Institute/NASA Ames Research Center\label{SETI}
         \and
         Department of Multi-Disciplinary Sciences, Graduate School of Arts and Sciences, The University of Tokyo, 3-8-1 Komaba, Meguro, Tokyo 153-8902, Japan\label{Dep_multi_tokio}
         \and
         Institute of Astronomy and Astrophysics, Academia Sinica, P.O. Box 23-141, Taipei 10617, Taiwan, R.O.C.\label{Taiwan_InstituteAA}
         \and
         Department of Astrophysics, National Taiwan University, Taipei 10617, Taiwan, R.O.C.\label{Taiwan_DepartmentA}
         \and
         NASA Ames Research Center, Moffett Field, CA 94035, USA\label{NASA_Ames}
         \and
         Department of Aeronautics and Astronautics, Massachusetts Institute of Technology, Cambridge, MA 02139, USA\label{MIT_Aeronautics}
         \and
         Department of Earth, Atmospheric, and Planetary Sciences, Massachusetts Institute of Technology, 77 Massachusetts Avenue, Cambridge, MA 02139, USA\label{MIT_EAPS}
         \and
         Department of Astrophysical Sciences, Princeton University, Princeton, NJ 08544, USA\label{Princeton}
         }

   \date{Received September 15, 1996; accepted March 16, 1997}

  \abstract
   {}
   {We report the discovery and validation of two TESS exoplanets orbiting faint M dwarfs: \toicuatrob and \toidosb.}
   {We have jointly analyzed space (TESS mission) and ground based (MuSCAT2, MuSCAT3 and SINISTRO instruments) lightcurves using our multi-color photometry transit analysis pipeline. This allowed us to compute contamination limits for both candidates and validate them as planet-sized companions.}
   {We found \toicuatrob to be a sub-Neptune-sized planet ($R_{p}=2.82^{+0.65}_{-0.63}~\rm R_{\oplus}$) and \toidosb to be a super-Earth-sized planet ($R_{p}=2.04^{+0.50}_{-0.54}~\rm R_{\oplus}$). Furthermore, we obtained that \toicuatrob, with a short orbital period of $1.15890^{+0.00002}_{-0.00001}~\rm days$, lies within the Neptune desert and is in fact the largest nearly ultra-short period planet around an M dwarf known to date.}
   {These results make \toicuatrob rare among the currently known exoplanet population around M dwarf stars, and an especially interesting target for spectroscopic follow-up and future studies of planet formation and evolution.}

   \keywords{Neptune desert -- Sub-Neptunes --
                Super-Earths --
                Stars: individual: \toicuatrob --
                Stars: individual: \toidosb --
                Planet and satellites: general --
                Methods: transits
               }

   \maketitle
%

\section{Introduction}

   Since the beginning of its observations in 2018, the \emph{Transiting Exoplanet Survey Satellite} \citep[TESS,][]{TESS} has provided nearly $5500$ objects of interest and $204$ confirmed planets\footnote{From the \texttt{NASA Exoplanet Archive}}. Although many of these objects of interest may be consistent with a planetary-like transit signal, not all of them have a planetary nature. Several astrophysical objects -- such as a brown dwarf or a low-mass star transiting a binary companion, a grazing binary stellar system, or a pair of blended binaries -- are able to mimic the signal of a planetary transit. Although these systems would produce deep eclipses, we could be observe their photometry diluted by a bright neighbor star \citep{Cameron2012,Ciardi_2015}. Therefore, the nature of each object of interest needs to be determined by ground-based follow-up observations, which play a supporting role by being able to confirm whether the candidate is a planet or not.

   Moreover, although the most reliable method to confirm a planet candidate is the mass determination through radial velocity (RV) measurements, this procedure is very difficult for those candidates orbiting a faint, active, or fast rotating star. Hence, to determine the nature of these candidates it is necessary to use other methods to validate them as planets.
   
   In this context, ground-based multicolor transit photometry is a useful method to validate planet candidates \citep{Drake2003,Tingley2004,Parviainen2019,Parviainen2020,Parviainen2021,Fukui_2022}. It allows to account for the light contamination from unresolved sources and estimate the uncontaminated radius ratio of the transiting candidate. With an estimate of the stellar radius, combined to the uncontaminated radius ratio, the radius of the candidate can be obtained. Consequently, if the radius of the candidate is significantly below the theoretical radius limit of a brown dwarf, the candidate can be validated as a planet \citep{Parviainen2020}.
   
   Here we use the approach of \cite{Parviainen2020} to validate the substellar nature of two TESS objects of interest orbiting faint ($V=13.369$ mag and $V=15.180$ mag) M dwarfs: the super-Earth-sized \toidosb and the sub-Neptune-sized \toicuatrob. We find \toicuatrob to be a rare target which lays on the Neptune desert.
   
   Our analysis is performed over space-based TESS photometry, ground-based datasets of multicolor photometry in \textit{$g$}, \textit{$r$}, \textit{$i$} and \textit{$z_{s}$} bands obtained through MuSCAT2 and MuSCAT3 multicolor imagers and complementary ground-based single-passband photometry from SINISTRO camera. We also obtained low-resolution optical spectra for the stellar characterization. In addition, high angular resolution observations are used to visually discard a binary stellar companion.

   In Section~\ref{sec: Tess_photometry} and Section~\ref{sec: Ground_Observations} we describe the observations used in our study. In Section~\ref{sec: lightcurve_analysis} we explain the methodology followed for the lightcurve analysis and the validation procedure. In Section~\ref{sec: Results}, we present and discuss our results, which confirm the planetary nature of \toidosb and \toicuatrob. Finally, we conclude our study in Section~\ref{sec: Conclusions}.


\section{TESS photometry}
\label{sec: Tess_photometry}

\toicuatro (TIC 126606859) was observed by TESS \citep{TESS} with two-minute cadence during Sector 41 during 29 days (from UTC 2021 July 23 to UTC 2021 August 20) during Cycle 4, obtaining 22 full transits in total. Full frame image (FFI) observations are available also during Sector 15, but they have not been used here. In this case, a transit signal with a 1.159 days orbital period and S/N = 9.5 was identified in the TESS Science Processing Operations Center \citep[SPOC,][]{Jenkins_2016} transiting planet search \citep{Jenkins_2002,Jenkins_2010,Jenkins20} of the two-minute data from Sector 41. The threshold crossing event was promoted to TESS Object of Interest (TOI) planet candidate status and designated TOI 4479.01 \citep{Guerrero_2021} based on a SPOC data validation report \citep{Twicken18_SPOC,Li_2019} showing clean transiting planet model fit and diagnostic test results.

\toidos (TIC 321669174) was observed by TESS with two-minute cadence during Sectors 14, 17, 20, 21, 24, 25, 26, 40, 41 and 47 (a total duration of 196 days during Cycle 2 and 59 days during Cycle 4), obtaining 22 full transits in total. A transit signal with a 10.504 days orbital period and S/N = 9.0 was identified in the TESS SPOC of the combined two-minute data from Sectors 14, 17, 20, and 21. The threshold crossing event was promoted to TOI planet candidate status and designated TOI 2081.01. 

\begin{figure}[h!]
    \centering
    \includegraphics[width=\columnwidth]{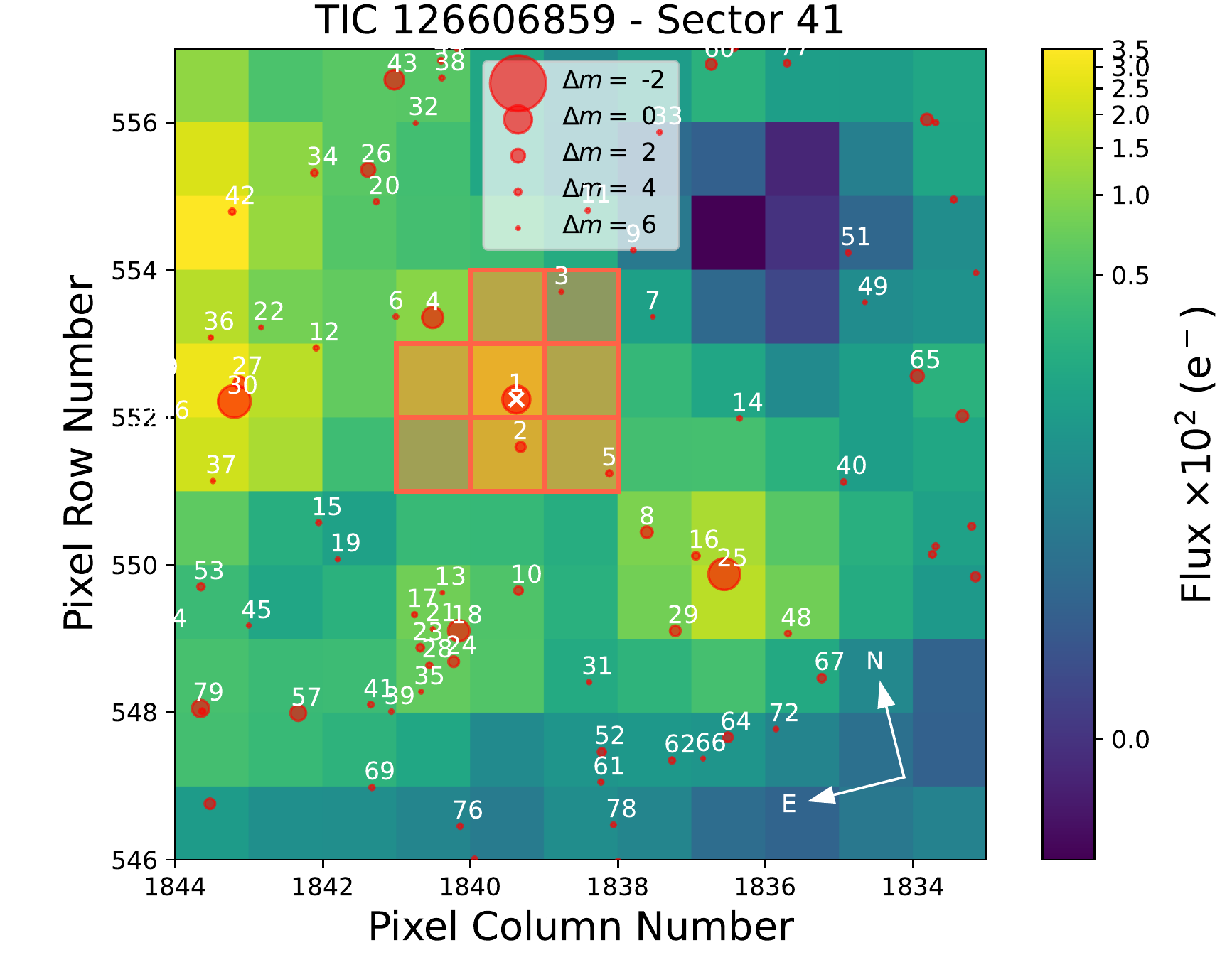}
    \includegraphics[width=\columnwidth]{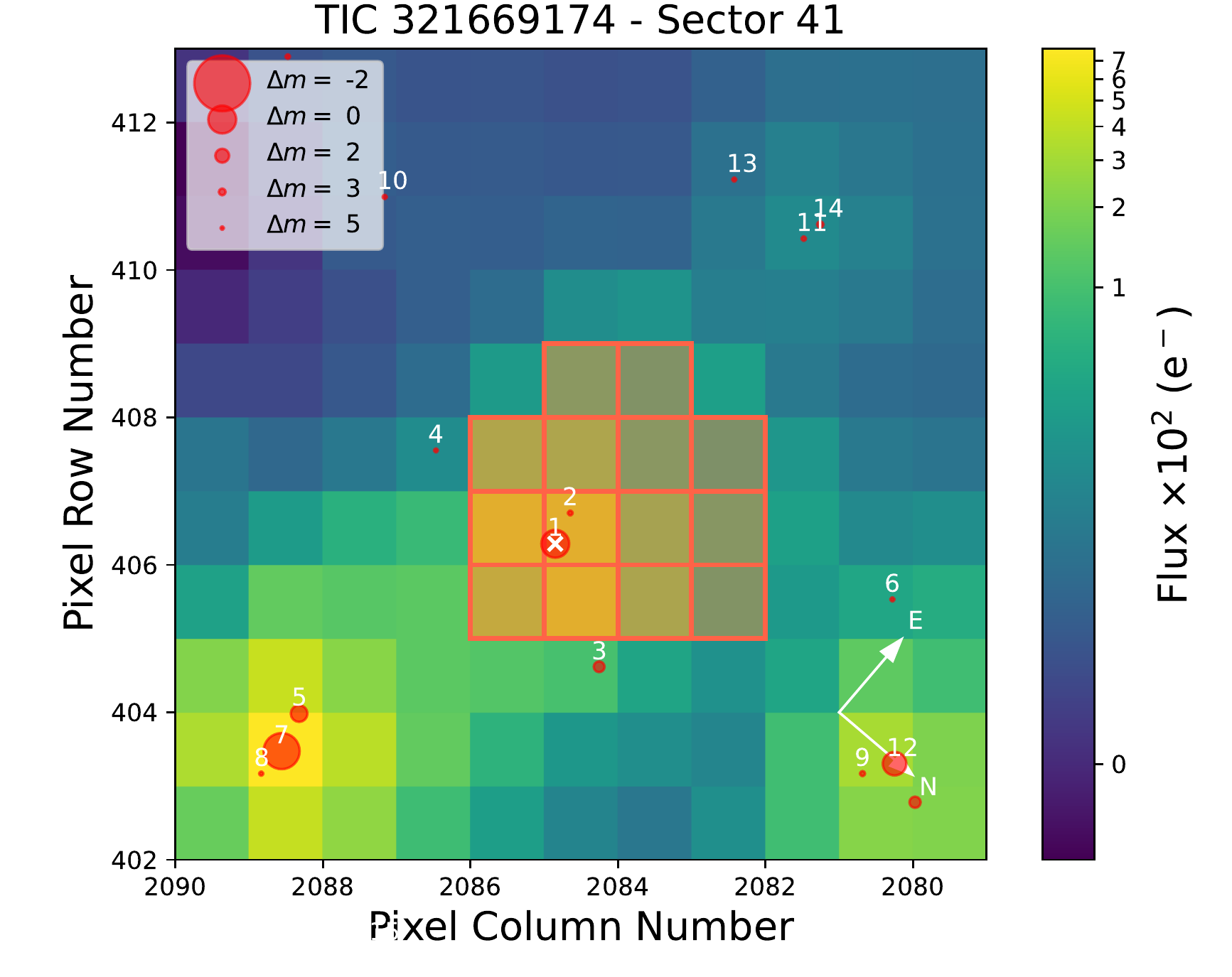}
    \caption{TESS target pixel file images of \toicuatro (top) and \toidos (bottom) observed in Sector 41. The red circles show the sources in the field identified by the \emph{Gaia DR2} catalogue \citep{GAIA_DR2} with scaled magnitudes. The position of the targets are indicated by white crosses and the mosaic of orange squares show the mask used by the pipeline to extract photometry. These plots were made using \texttt{tpfplotter} \citep{Aller2020_tpfplotter}.}
    \label{fig:tpf_plots}
\end{figure}

The TESS images around the position of \toicuatro and \toidos in Sector 41 are shown in Figure~\ref{fig:tpf_plots}. The TESS images in the rest of sectors where \toidos was observed are shown in Figure~\ref{fig:TPF_2081_mosaic}.

\section{Ground-based Follow-up Observations}
\label{sec: Ground_Observations}

\subsection{MuSCAT2 photometry}
We observed a full transit of \toicuatrob on UTC 2021 October 17 and a full transit of \toidosb on UTC 2020 July 22 with the MuSCAT2 instrument \citep{Narita2019_MuSCAT2}, mounted in Telescopio Carlos Sánchez (TCS) at the Teide Observatory, Spain. MuSCAT2 is a multicolor imager capable of performing simultaneous photometry in the \textit{g}, \textit{r}, \textit{i} and \textit{$z_{s}$} photometric bands using 4 independent CCDs. The exposure times used for \toicuatro and \toidos observations were set independently for each CCD (\textit{g}, \textit{r}, \textit{i}, \textit{$z_{s}$}): (15, 30, 25, 20) and (10, 25, 25, 15) seconds, respectively. As the objects are red, the exposure times are shorter in redder filters to avoid saturation. The g filter is an exception: since the flux there is small, we set a short integration time and use this channel images to auto-guide, for which we want an exposure time typically $<15$s.

\begin{figure*}[h!]
    \centering
    \centerline{
    \includegraphics[width=0.5\textwidth]{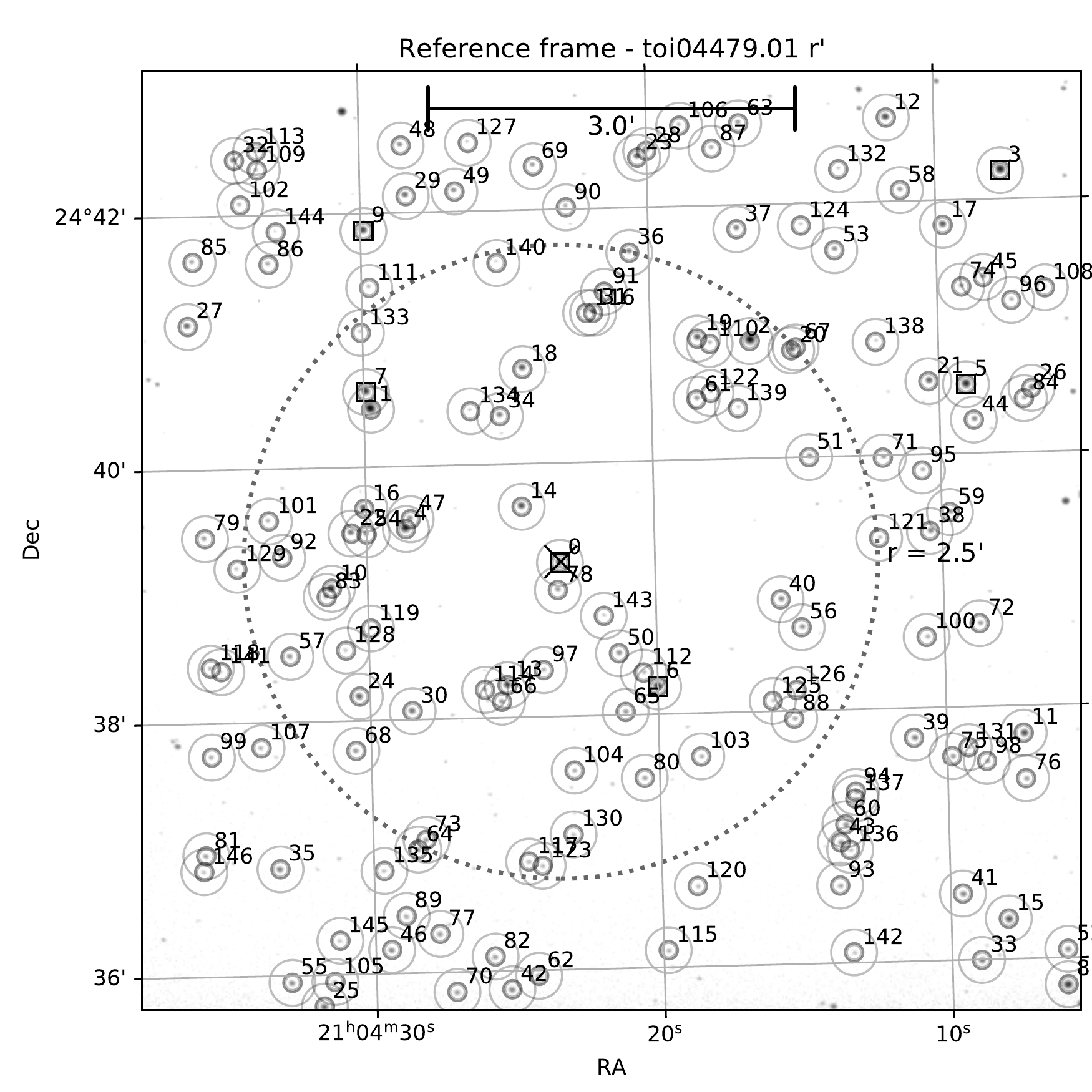}
    \includegraphics[width=0.5\textwidth]{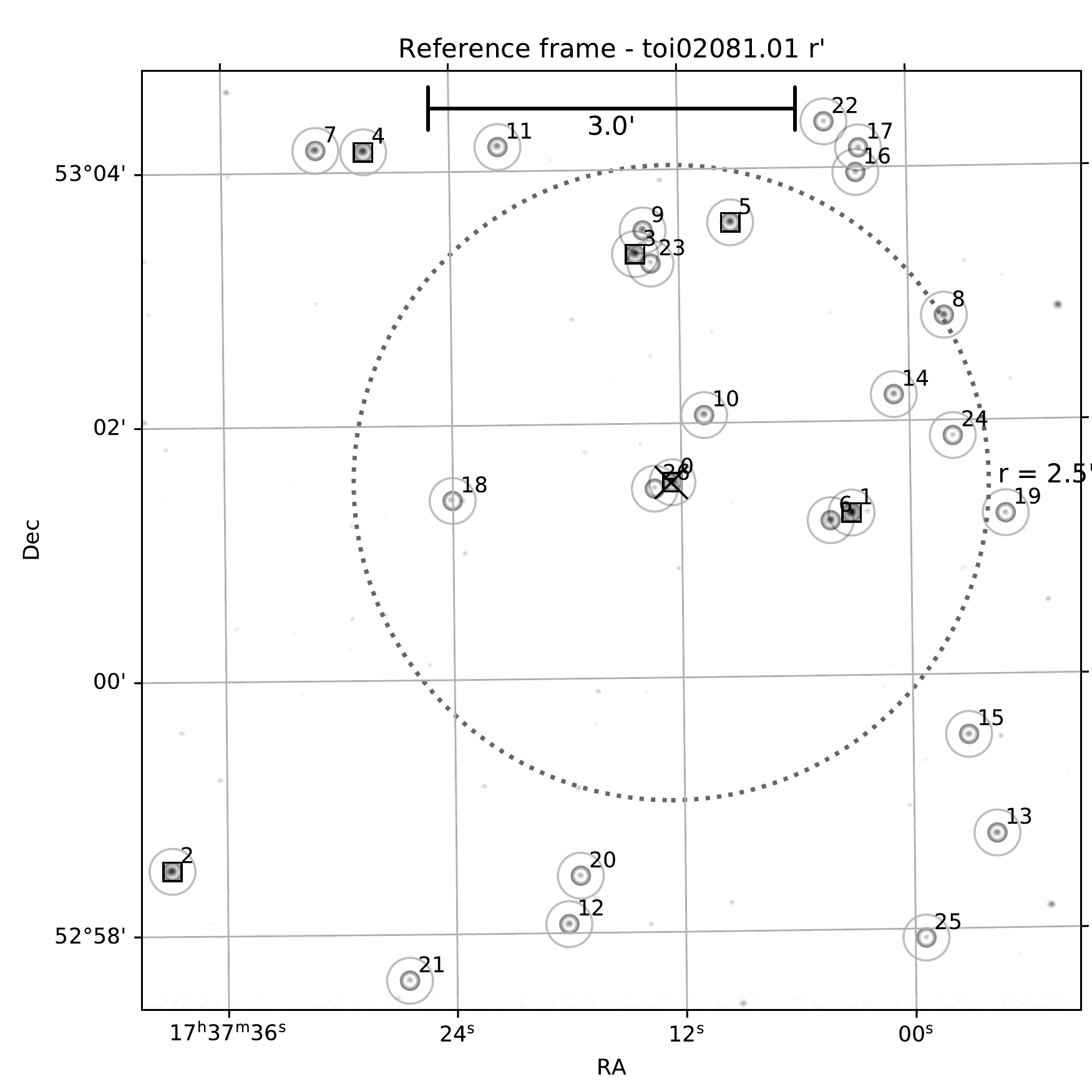}}
    \caption{MuSCAT2 field in the \textit{r} band during the \toicuatro (left) and \toidos (right) observations. The cross indicates the position of the target, and the dotted circle marks the $2.5'$-radius region centered around it.}
    \label{fig:M2_FOV_TOI2081b}
\end{figure*}

A dedicated MuSCAT2 pipeline, described by \cite{Parviainen2020}, was used to perform the data reduction and to extract the photometry. The pipeline performs aperture photometry for a set of comparison stars and aperture sizes (see Figure~\ref{fig:M2_FOV_TOI2081b}). The final relative light curves are obtained through global optimization of a model, which aims to find the optimal comparison stars and aperture sizes while the transit and baseline variations are simultaneously modeled using a linear combination of covariates.

\subsection{MuSCAT3 photometry}

A full transit of \toicuatrob was observed simultaneously in Sloan $g$, $r$, $i$, and Pan-STARRS $z$-short bands on UTC 2021 October 21 using the Las Cumbres Observatory Global Telescope \citep[LCOGT;][]{Brown:2013} 2\,m Faulkes Telescope North at Haleakala Observatory on Maui, Hawai'i. The telescope is equipped with the MuSCAT3 multi-band imager \citep{Narita:2020}. We used the {\tt TESS Transit Finder} to schedule our transit observations. The images were calibrated using the standard LCOGT BANZAI pipeline, and the differential photometric data were extracted in each band using {\tt AstroImageJ} \citep{Collins:2017} with circular apertures having radius $2\farcs7$. The apertures exclude virtually all of the flux from the nearest Gaia EDR3 \citep{GAIAEDR3} neighbor (TIC 1951208113) $8\farcs7$ east of the target. The transit was detected on-target in all four filter bands.

\subsection{LCOGT 1\,m photometry}

We observed a full transit of \toicuatrob from the Las Cumbres Observatory Global Telescope LCOGT 1.0\,m network on UTC 2021 October 11 in Sloan $i'$ band. We used the {\tt TESS Transit Finder} \citep{TTF} to schedule our transit observations. The 1\,m telescopes are equipped with $4096\times4096$ SINISTRO cameras having an image scale of $0\farcs389$ per pixel, resulting in a $26\arcmin\times26\arcmin$ field of view. The images were calibrated by the standard LCOGT {\tt BANZAI} pipeline \citep{McCully:2018}. The differential photometric data were extracted using {\tt AstroImageJ} \citep{Collins:2017} with circular photometric apertures having radius $4\farcs7$. The target star aperture excludes most of the flux of the nearest Gaia EDR3 and TESS Input Catalog neighbor (TIC 1951208113) $8\farcs7$ east of the target. The transit was detected on-target.

\subsection{High resolution imaging of \toidos}

\begin{figure}[h!]
    \centering
    
    \includegraphics[width=\columnwidth]{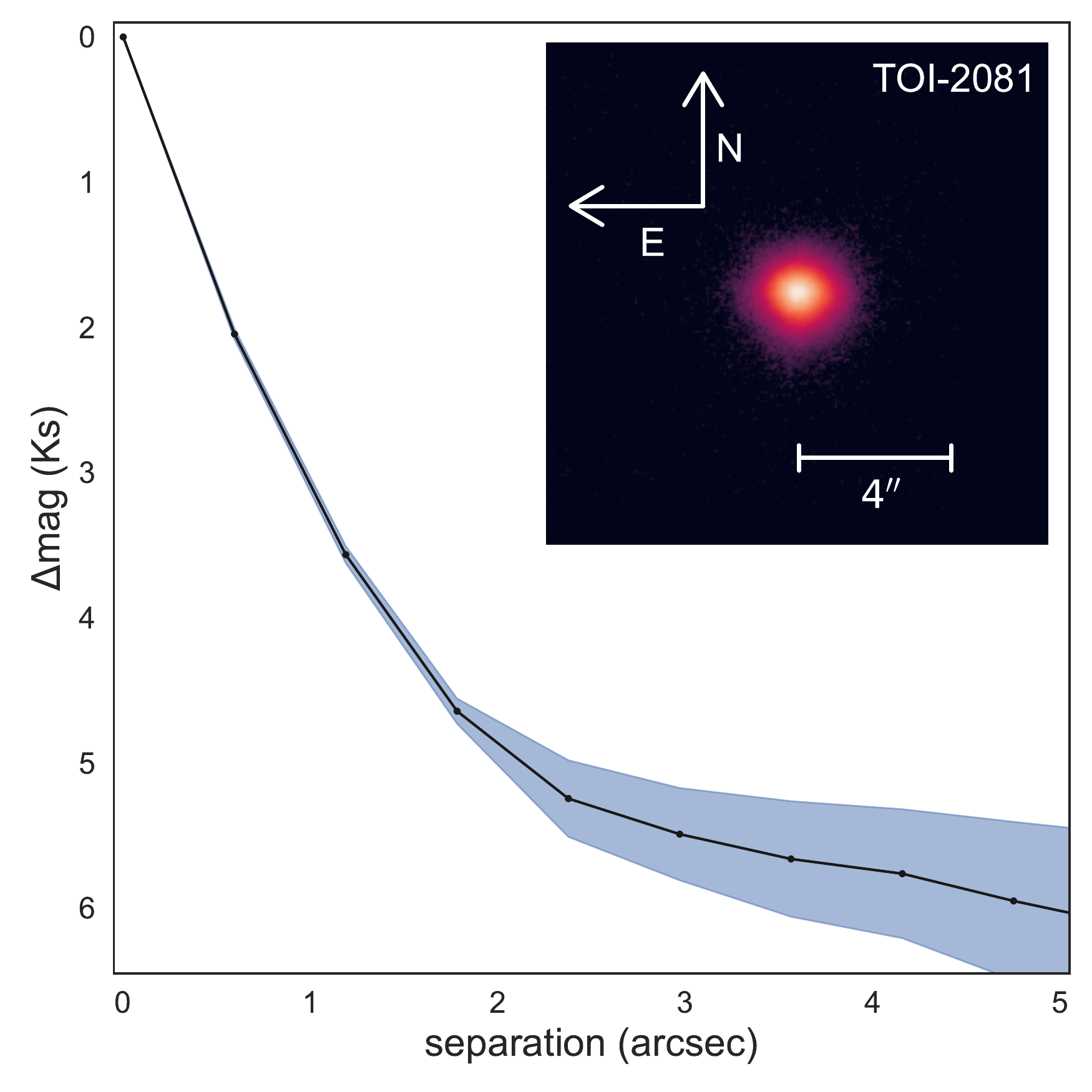}
    \includegraphics[width=\columnwidth]{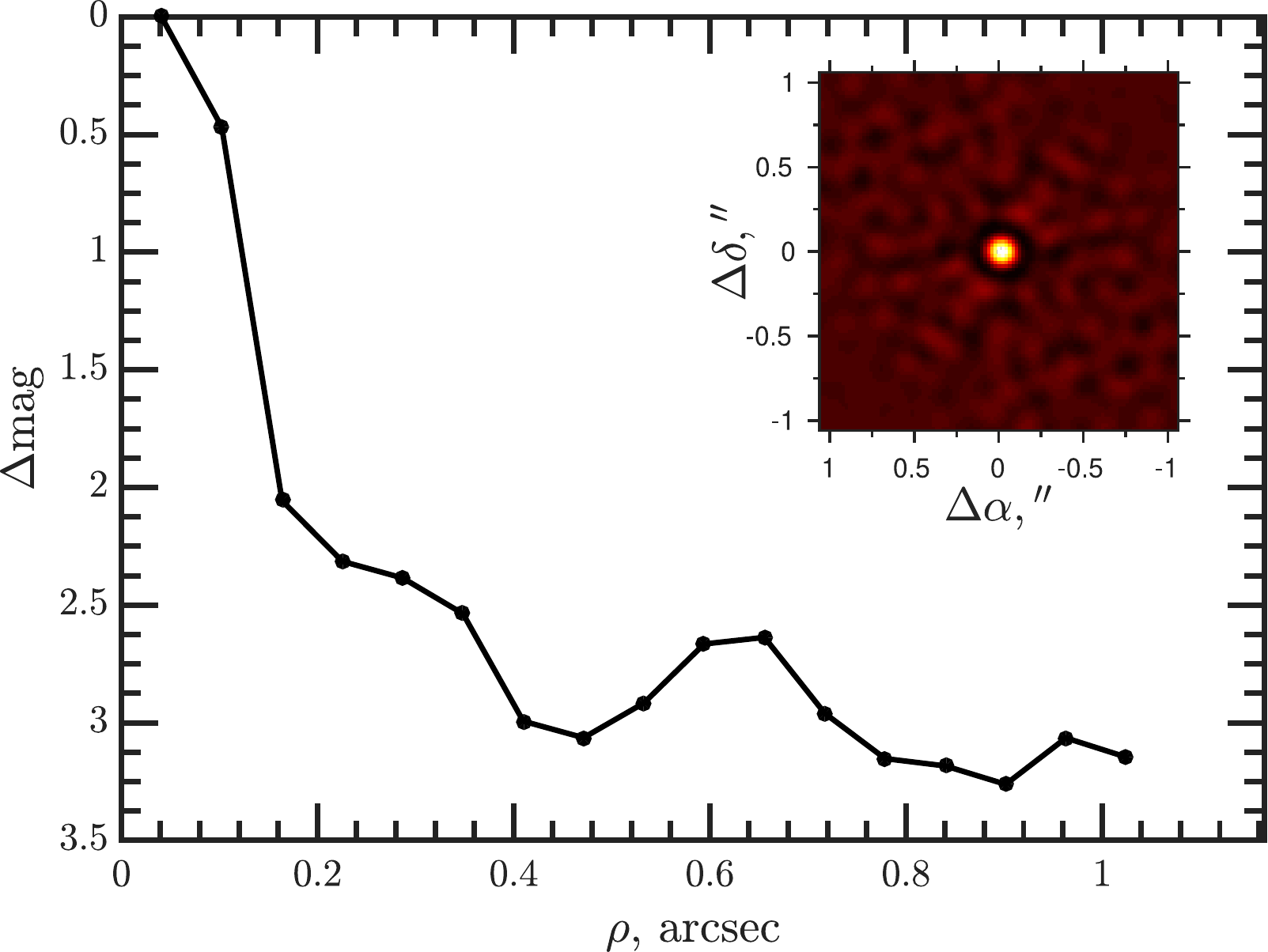}

    \caption{Top: Adaptive optics images of TIC 321669174 taken with the ShARCS camera on the Shane 3-meter telescope at Lick Observatory. For each image, we also present a contrast curve generated by calculating the median values (solid lines) and root-mean-square errors (blue, shaded regions) in annuli centered on each target, where the bin width of each annulus is equal to the full width at half max of the point spread function. The observation rule out the presence of a possible nearby contaminant companion. Bottom: High angular resolution speckle imaging of \toidos in I filter using SAI-2.5m telescope. The observation rules out the presence of a possible nearby contaminant companion.The structure in the lower contrast curve is due to the first derivative discontinuity in the frequency mask that is used to compute the autocorrelation. This discontinuity generates rings at characteristic distances from the center, which have a minor effect in the detection limit.}
    \label{fig:DI_Shane}
\end{figure}

We observed \toidos on UTC 2021 March 29 using the ShARCS camera on the Shane 3-meter telescope at Lick Observatory \citep{2012SPIE.8447E..3GK, 2014SPIE.9148E..05G, 2014SPIE.9148E..3AM}. Observations were taken with the Shane adaptive optics system in natural guide star mode in order to search for nearby, unresolved stellar companions. We collected one sequence of observations using a $Ks$ filter ($\lambda_0 = 2.150$ $\mu$m, $\Delta \lambda = 0.320$ $\mu$m) and reduced the data using the publicly available \texttt{SImMER} pipeline \citep{2020AJ....160..287S}.\footnote{\url{https://github.com/arjunsavel/SImMER}} Our reduced images and corresponding contrast curves are shown in the top panel of Figure~\ref{fig:DI_Shane}. We find no nearby stellar companions within our detection limits.

We also observed \toidos on UTC 2021 March 03 with the Speckle Polarimeter \citep[SPP,][]{Safonov2017} on the 2.5~m telescope at the Caucasian Observatory of Sternberg Astronomical Institute (SAI) of Lomonosov Moscow State University. SPP uses Electron Multiplying CCD Andor iXon 897 as a detector. The atmospheric dispersion compensator allowed observation of this relatively faint target through the wide-band $I_c$ filter. The power spectrum was estimated from 4000 frames with 30 ms exposure. The detector has a pixel scale of $20.6$ mas pixel$^{-1}$, and the angular resolution was 89 mas. We did not detect any stellar companions brighter than $\Delta I_C=2.4$ and $3.1$ at $\rho=0\farcs25$ and $1\farcs0$, respectively, where $\rho$ is the separation between the source and the potential companion.

Nearby faint companions of the host star may remain undetected through seeing-limited photometry, but could contribute to a contamination of the flux of the target and lead to wrong estimations of the planetary radius. However, the Shane-AO and SAI-Speckle (Figure~\ref{fig:DI_Shane}) observations allow us to rule out this scenario for \toidos in $K_{s}$ and $I$ bands. We do not detect any nearby visual companion and \toidos seems an isolated star in the data.

	\begin{table}[h!]    
		\caption{\toicuatro and \toidos identifiers, coordinates, magnitudes and stellar parameters.}
		\centering
		\begin{tabular*}{\columnwidth}{@{\extracolsep{\fill}} lll}
			\hline\hline
			\emph{Main identifiers} & \toicuatro & \toidos    \\
			\hline     
			TIC   & {\footnotesize 126606859} & {\footnotesize 321669174} \\
			2MASS & {\small J21042315+2439153} &  {\small J17371272+5301326}  \\ 
			WISE  & {\small J210423.23+243913.8} & {\small J173712.60+530132.2} \\
			\\
			\multicolumn{3}{l}{\emph{Equatorial coordinates}}     \\
			\hline            
			RA \,(J2000) & $21^h\,04^m\,23\fs27$ & $17^h\,37^m\,12\fs54$\\
			Dec (J2000)  & $24\degr\,39\arcmin\,13\farcs23$ & $53\degr\,01\arcmin\,32\farcs04$\\
			\\     
			\multicolumn{3}{l}{\emph{Magnitudes}} \\
			\hline              
			\centering
			TESS & $12.9374 \pm 0.0075$ & $11.642 \pm 0.007$\\
			$V$  & $15.2 \pm 0.2$ & $13.369 \pm 0.035$\\
			Gaia DR2 & $14.1309 \pm 0.0005$ & $12.6594 \pm 0.0003$\\
			$J$  & $11.44 \pm 0.02$ & $10.36 \pm 0.02$\\
			$H$  & $10.85 \pm 0.02$ & $9.75 \pm 0.03$\\
			$K$  & $10.65 \pm 0.02$ & $9.52 \pm 0.02$\\
			\\     
			\multicolumn{3}{l}{\emph{Stellar parameters$^{1}$}} \\
			\hline              
			\centering
			Spectral Type & $\rm M3.0 \pm 0.5$ & $\rm M1.0 \pm 0.5$\\
			$M_\star$ [$\rm M_{\odot}$] & $ 0.452 \pm 0.090$ & $ 0.540 \pm 0.080$\\
            $R_\star$ [$\rm R_{\odot}$] & $0.451 \pm 0.085$ & $0.534 \pm 0.080$\\
            $L_\star$ [$\rm L_{\odot}$] & $0.02487 \pm 0.00015$ & $0.04587 \pm 0.00023$\\
            $\log{g}$ [dex] & $\ge$4.5 & $\ge$ 4.5\\
            $T_{\rm eff}$ [K] & $3400 \pm 100$ & $3800 \pm 100$\\
            $[{\rm Fe/H}]$ [dex] & $\ge$ 0.0 & $\ge -0.5$\\
            Parallax [mas]$^{2}$ & $12.41 \pm 0.02$ & $16.05 \pm 0.01$\\
            Distance [pc]$^{2}$ & $80.6 \pm 0.1$ & $62.31 \pm 0.05$\\
			\hline
		\end{tabular*}
		\tablefoot{$^{1}$ Derived from ALFOSC spectroscopy and alanysis of the stellar SED. $^{2}$ Gaia EDR3}
		\label{tab:stellar_parameters}  
	\end{table}

\subsection{ALFOSC spectroscopy}
\label{sec: alfosc_spectroscopy}

\toidos and \toicuatro were spectroscopically observed with the Alhambra Faint Object Spectrograph and Camera (ALFOSC), mounted on the Cassegrain focus of the 2.5-m Nordic Optical Telescope (NOT) on the Observatorio del Roque de los Muchachos (La Palma, Spain), on 2022 April 8 and 10 UT, respectively. ALFOSC is equipped with a monolithic 2048\,$\times$\,2048 E2V detector that has a pixel size of 0\farcs2138 on the sky. On both nights, we used a long-slit with a slit-width of 1\farcs0 and the grating number 5. This instrumental configuration yields low-resolution spectra covering the optical wavelength interval from 500 through 1050 nm with a nominal dispersion of 3.38 \AA\,pixel$^{-1}$ (or resolving power $R = 610$ at 725 nm). Fringing is, however, strong ($\ge 4 \%$) redwards of $\approx$900 nm; therefore, we discard all data at longer wavelengths. Two exposures of 900 s each were acquired for \toidos and \toicuatro at an air mass of 1.10 and 1.58, respectively. Together with the main targets, we also observed the spectroscopic standard star BD+26\,2606 at different air masses for a proper correction of the instrumental response and telluric absorption. Three exposures of 15 s each were acquired on each night for the stardard star. BD+26\,2606 is an early-type star with known fluxes published in \citet{oke90}. All observations were acquired at parallactic angle to minimize light losses on the slit. We windowed the ALFOSC detector along the spatial axis (perpendicular to the dispersion axis) to a size of 500 pixels.

\begin{figure}[h!]
    \centering
    \includegraphics[width=\columnwidth]{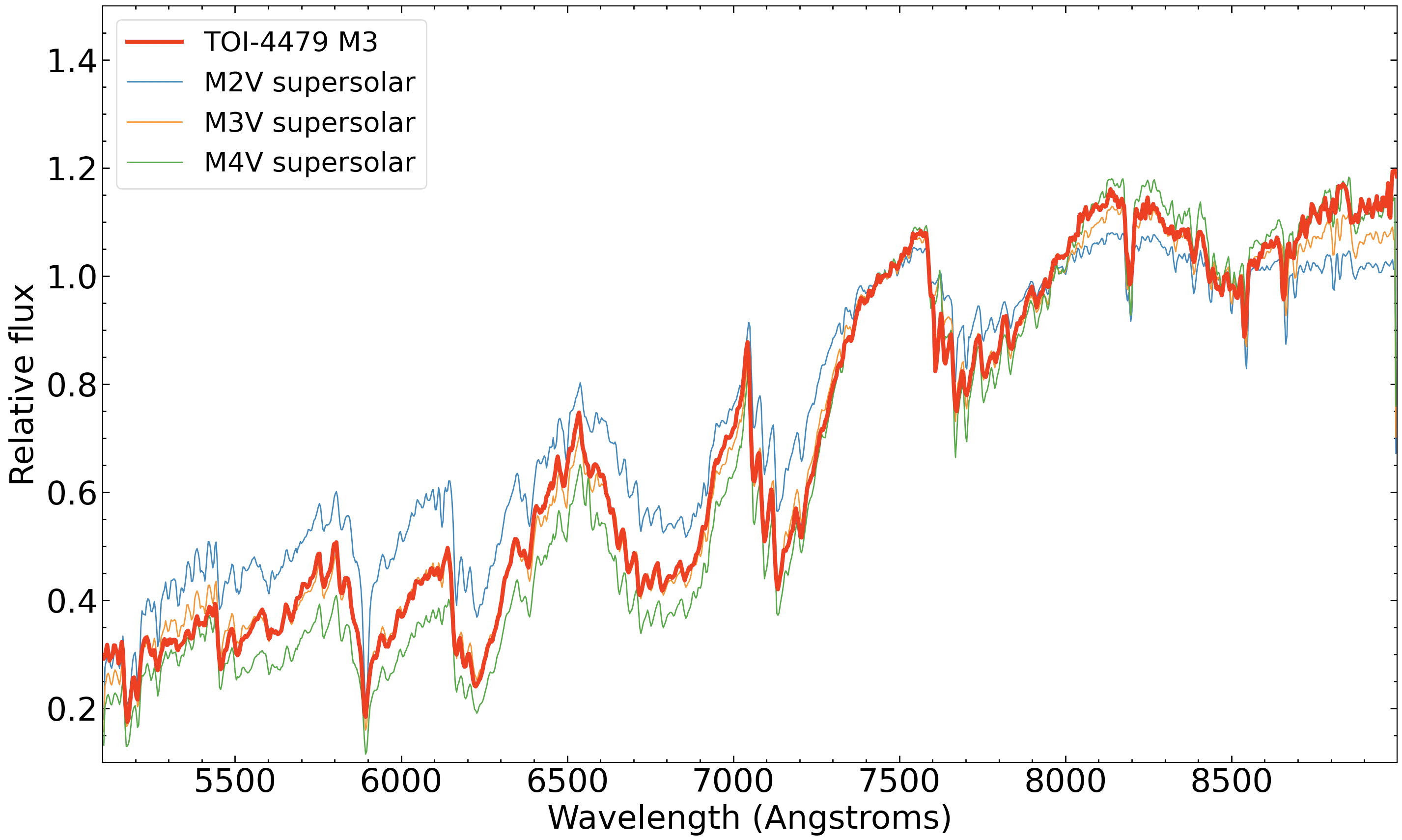}
    \includegraphics[width=\columnwidth]{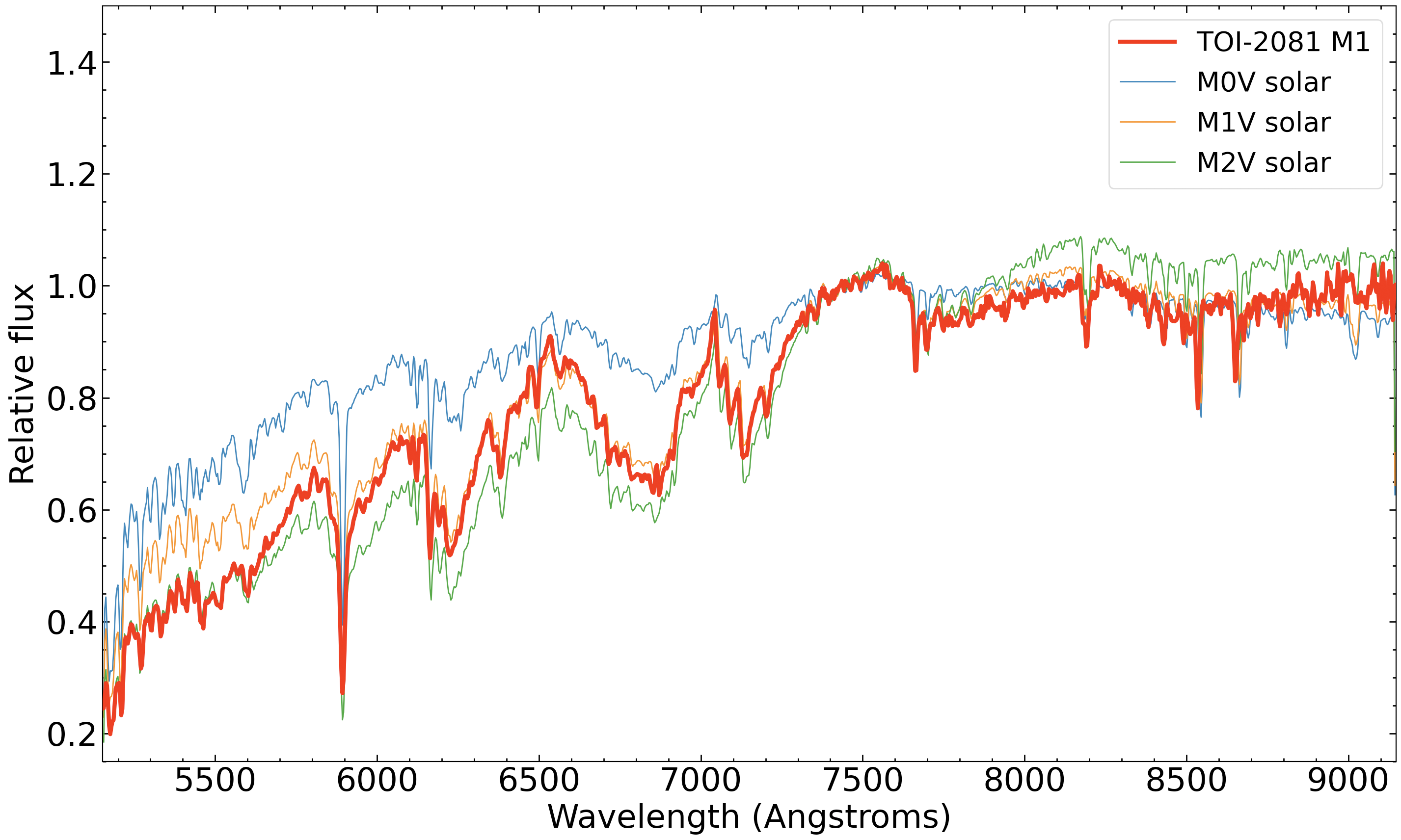}
    \caption{ALFOSC spectra (red) of \toicuatro (top panel) and \toidos (bottom panel) are shown in comparison with solar metallicity templates of known spectral type from the library of \citet{kesseli17}.}
    \label{fig:alfosc_spectra}
\end{figure}

The ALFOSC spectra of the targets and the standard star were reduced and optimally extracted following standard steps within the IRAF environment \citep{tody93}. First, we removed the detector bias at the same time we subtract "the sky" contribution using the region on both sides of the spectral trace of the stars. The spectra were calibrated in wavelength with a precision of about 1.5\,\AA~using observations of He$+$Ne lamps acquired immediately after observing the main targets. The ALFOSC spectra of \toidos and \toicuatro were corrected for instrumental response using the observations of the standard star. The final step was to mask out the hydrogen lines intrinsic to the standard star and to normalize its spectra to the continuum for division with the target's spectrum (in this way, we removed
the Earth's telluric lines from the spectrum). The telluric-free spectra of \toidos and \toicuatro are shown in Figure~\ref{fig:alfosc_spectra}.

\section{Stellar parameters}
In Figure~\ref{fig:alfosc_spectra}, the ALFOSC spectra are compared to spectral standard stars from the library of empirical stellar spectra of \citet{kesseli17}. Our data are characterized by the presence of TiO absorption over the optical wavelengths, which is a signpost of M-spectral classification. Using templates of solar metallicity, we derived the following spectral types: M1.0 $\pm$ 0.5 (TOI-2081) and M3.0 $\pm$ 0.5 (TOI-4479). The ALFOSC spectra are well reproduced by the templates with no significant deviations. H$\alpha$ is not in emission in any of the two dwarfs. \citet{kesseli17} template spectra are binned by metallicity from $-$2.0 dex through $+$1.0 dex, and are separated into main-sequence (dwarf) stars and giant stars. We also compared the ALFOSC data to the sets of different metallicities and gravities finding that \toidos and \toicuatro are better described by high-gravity surfaces and atmospheric metallicity [Fe/H] $\ge$-0.5 and $\ge$0.0 dex, respectively, thus supporting the solar-to-metal-rich nature of both stars. At low resolution, we cannot better constrain the metallicity or surface gravity of the targets.

To derive the stellar mass and radius, we first built the stars' spectral energy distribution (SED) by combining the ALFOSC data and all available broad-band photometry from {\sl Gaia} DR3 \citep{GAIAEDR3}, the $JHK$ magnitudes from 2MASS \citep{2MASS}, and the $W1-W4$ magnitudes from {\sl WISE} \citep{WISE}. All these data were converted into the absolute fluxes by using the {\sl Gaia} trigonometric distances. The extension of the SEDs towards bluer and redder wavelengths was done with the BT-Settl model \citep{all12} that best reproduces the observations. The integration of the SEDs yields the bolometric luminosities provided in Table~\ref{tab:stellar_parameters}. We then employed the bolometric luminosities and the mass--radius--luminosity relations of \citet{Cifuentes20} to derive \toidos and \toicuatro stellar parameters, which are listed in Table~\ref{tab:stellar_parameters} and will be used in our analysis of the planetary systems.

\section{Lightcurve analysis}
\label{sec: lightcurve_analysis}

\subsection{Multicolor transit analysis}

The TESS\footnote{The TESS SPOC Presearch Data Conditioning Simple Aperture Photometry \citep[PDCDAP;][]{Stumpe_2012,Stumpe_2014,Smith_2012} lightcurves.} and ground-based lightcurves were analyzed individually and jointly following the procedure described in \cite{Parviainen2019, Parviainen2020, Parviainen2021}, which performs an exoplanet-orientated Bayesian parameters estimation \citep{Parviainen2018}. The multicolor analysis procedure can be summarized in the following steps:

\begin{enumerate}
    \item A flux model is generated to fit the light curves, accounting for both the transit signal and the systematic effects present in the time series.
    \item A noise model is defined to account for the stochastic variability in the data.
    \item The likelihood is obtained combining the flux model, the noise model and the observations.
    \item Finally, a Markov Chain Monte Carlo (MCMC) sampling is performed to obtain the joint parameter posterior distribution based on the priors defined from the model parameters.
\end{enumerate}

The pipeline used to perform the multicolor analysis makes use of \texttt{PHOENIX} \citep{Husser13} for physics-based contamination modeling, \texttt{LDTk} \citep{LDTk_Parviainen} for limb-darkening estimations, \texttt{PyTRANSIT} \citep{PyTRANSIT1}, which provides flux and noise models, and \texttt{emcee} \citep{emcee} to perform the MCMC sampling. 

We applied the multicolor analysis to the complete set of photometric data of both targets. The \toicuatro dataset is composed of 31 transits (22 TESS transits, one MuSCAT2 transit observed simultaneously in $g$, $r$, $i$, $z_{s}$ bands, one MuSCAT3 transit observed simultaneously in $g_{p}$, $r_{p}$, $i_{p}$, $z_{s}$ bands and one SINISTRO transit in $i_{p}$ band). The \toidos dataset is composed of 26 transits in total (22 TESS transits and one MuSCAT2 transit observed simultaneously in $g$, $r$, $i$, $z_{s}$ bands).

In our calculations we adopted the values for the stellar parameters shown in Table~\ref{tab:stellar_parameters} taken from the TESS Input Catalog \texttt{TICv8} \citep{Stassun2019_TIC}.

\begin{figure*}[ph]
    
    \begin{center}
        \centerline{\includegraphics[width=1.05\textwidth]{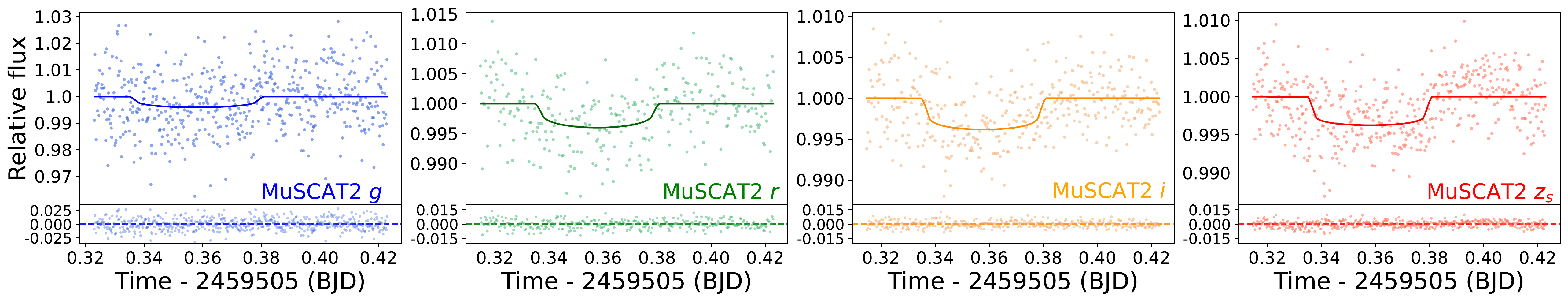}}
        \centerline{\includegraphics[width=1.05\textwidth]{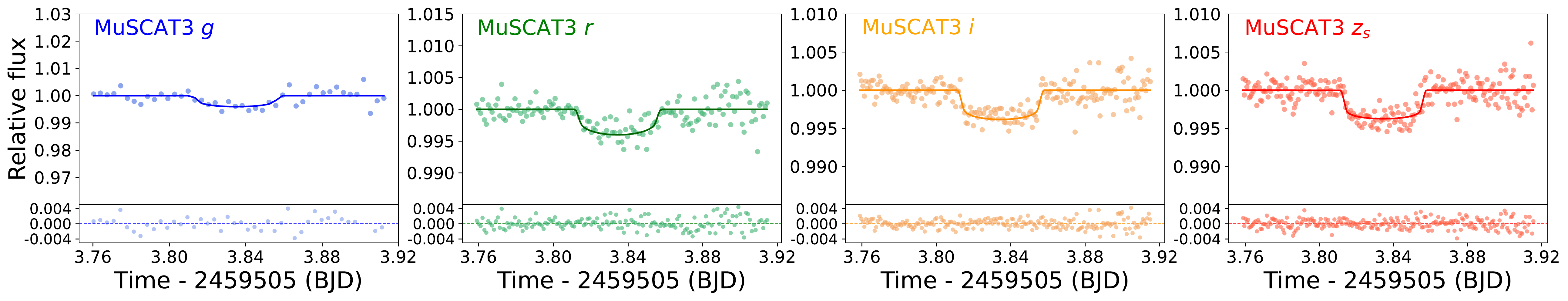}}
        \caption{MuSCAT2 (top) and MUSCAT3 (bottom) detrended lightcurves of \toicuatrob in \emph{g}, \emph{r}, \emph{i} and \emph{$z_{s}$} passbands. The dots show the MuSCAT2 and MUSCAT3 relative flux and the lines show the best lightcurve model from the MuSCAT2 pipeline for each band.}
            \label{fig:M2_LC}
    \end{center}
    
    \begin{center}
        \centerline{\includegraphics[width=1.05\textwidth]{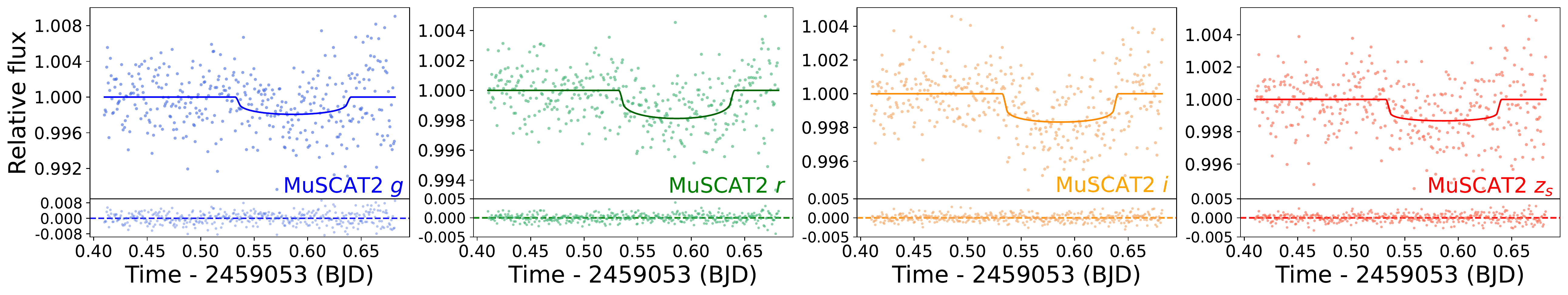}}
            \caption{MuSCAT2 detrended lightcurves of \toidosb in \emph{$g$}, \emph{$r$}, \emph{$i$} and \emph{$z_{s}$} passbands. The dots show the MuSCAT2 relative flux and the lines show the best lightcurve model.}
            \label{fig:M3_LC}
    \end{center}
    
    \begin{center}
        \centerline{\includegraphics[width=0.5\textwidth]{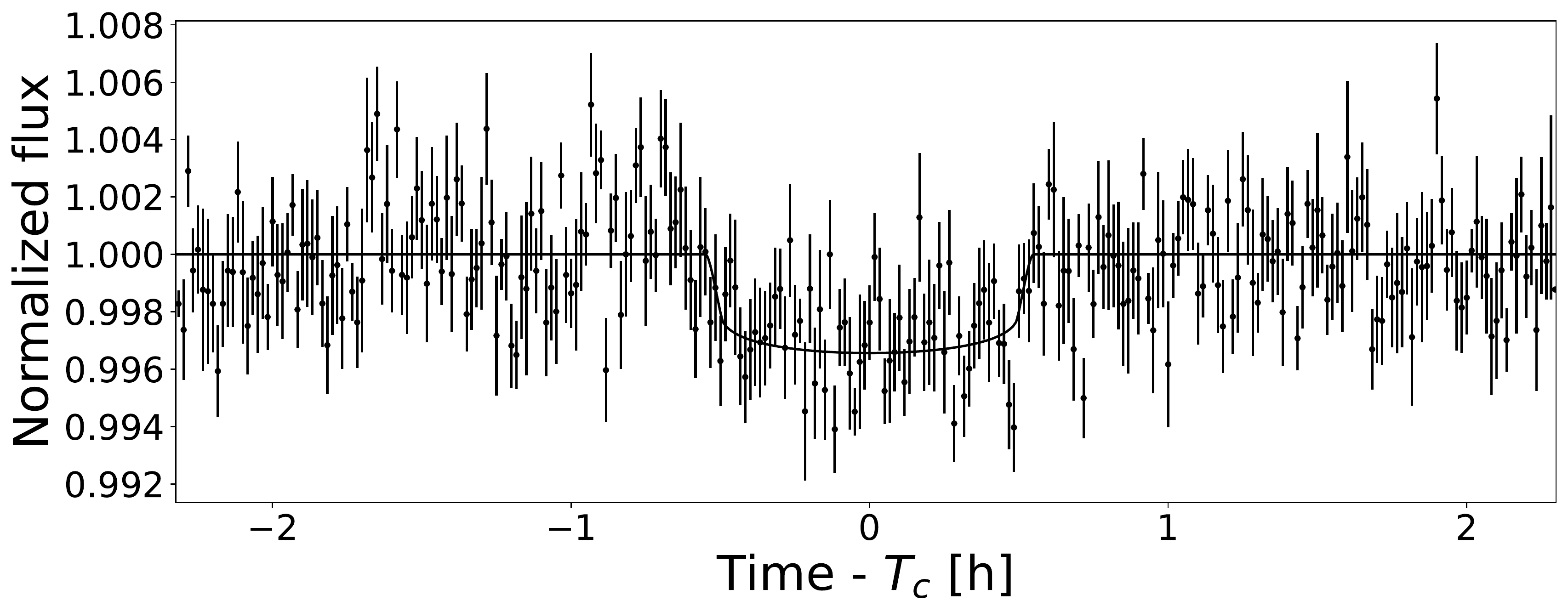}
        \includegraphics[width=0.48\textwidth]{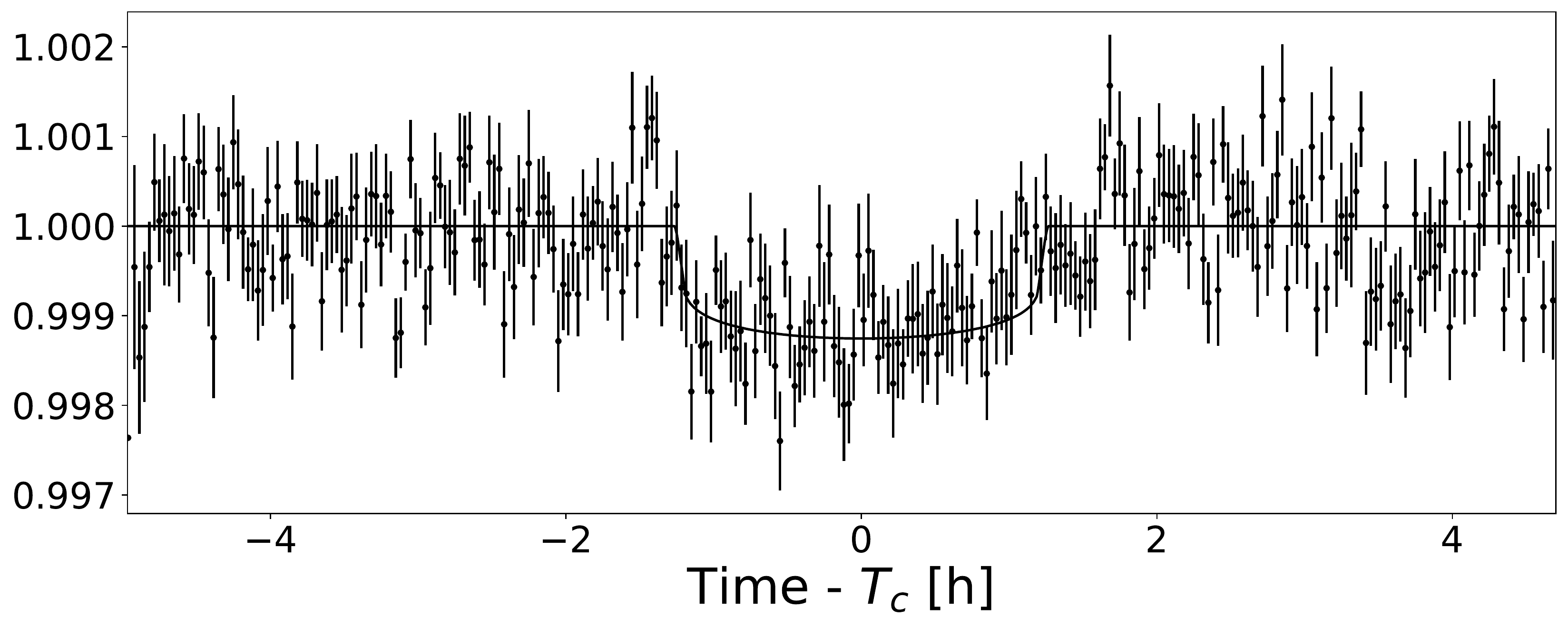}}
        \caption{TESS PDCSAP folded relative flux from the combination of all the transits and best lightcurve model of \toicuatrob (left) and \toidosb (right).}
        \label{fig:TESS_LCs}
    \end{center}

    \begin{center}
        \centerline{\includegraphics[width=0.5\textwidth]{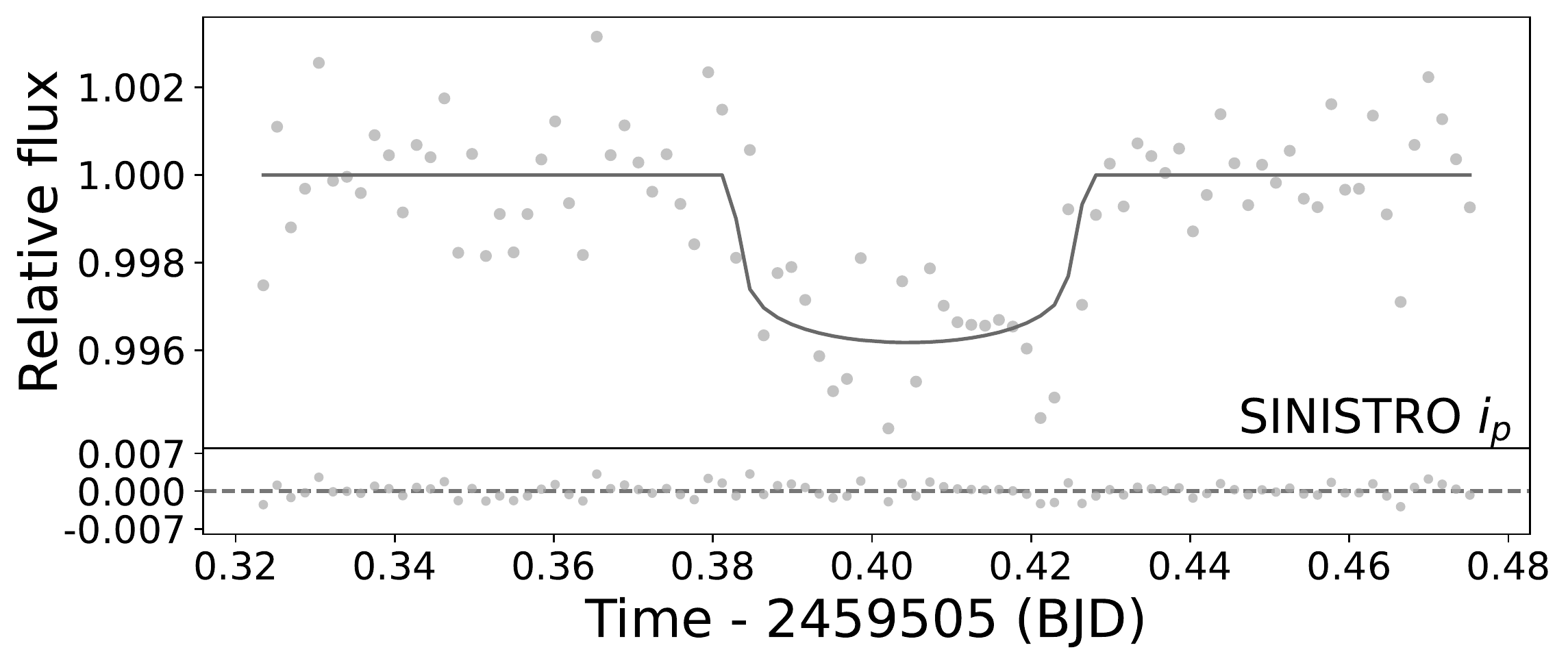}}
        \caption{SINISTRO detrended lightcurve of \toicuatrob in the \emph{$i_{p}$} passband. The dots show the SINISTRO relative flux and the line shows the best lightcurve model.}
        \label{fig:LCO_LC}
    \end{center}

\end{figure*}

\subsection{Contamination analysis}

The possible multiplicity of the system and the subsequent presence of an unresolved companion entails a flux contamination of the host star that may affect the observed transit depth and lead to erroneous parameters of the planetary system \citep{Daemgen09}. Single-passband photometry is not able to constrain such contamination because of the degeneracy with orbital geometry, limb darkening and planet-to-star radius ratio. Nevertheless, as described in \cite{Parviainen2019}, some effects of the flux contamination are color-dependent \citep{Rosenblatt1971,Drake2003,Tingley2004}, making multicolor photometry a valuable tool to constrain the degree of contamination and estimate the true planet-to-star radius ratio. On the one hand, color differences between the host star and the companion may lead to significant variations in the transit depth in different passbands. On the other hand, the transiting object produces a color-dependent signal, leaving a distinctive signature that relies either on the radius of the transiting object and the nature of the transiting object. Both effects allow to discriminate whether a transiting planet candidate is actually a planet or, conversely, a mimicked signal by flux contamination.

Our multicolor transit analysis pipeline accounts for the effects of flux contamination and estimates, among the system parameters, the posterior distributions of the apparent radius ratio, the true radius ratio (free of the contribution of the contaminant) and the effective temperature of the possible contaminant companion. By analyzing these three parameters, we are able to evaluate the flux contamination and to validate the nature of the companion.

\begin{figure*}[h!]
    
    \begin{center}
        \centerline{\includegraphics[width=\textwidth]{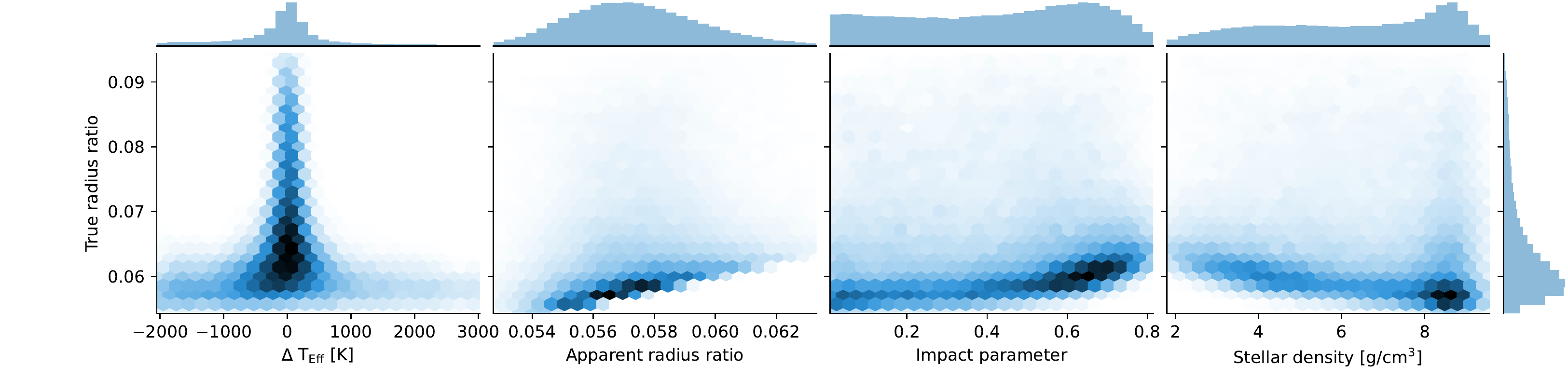}}
        \centerline{\includegraphics[width=\textwidth]{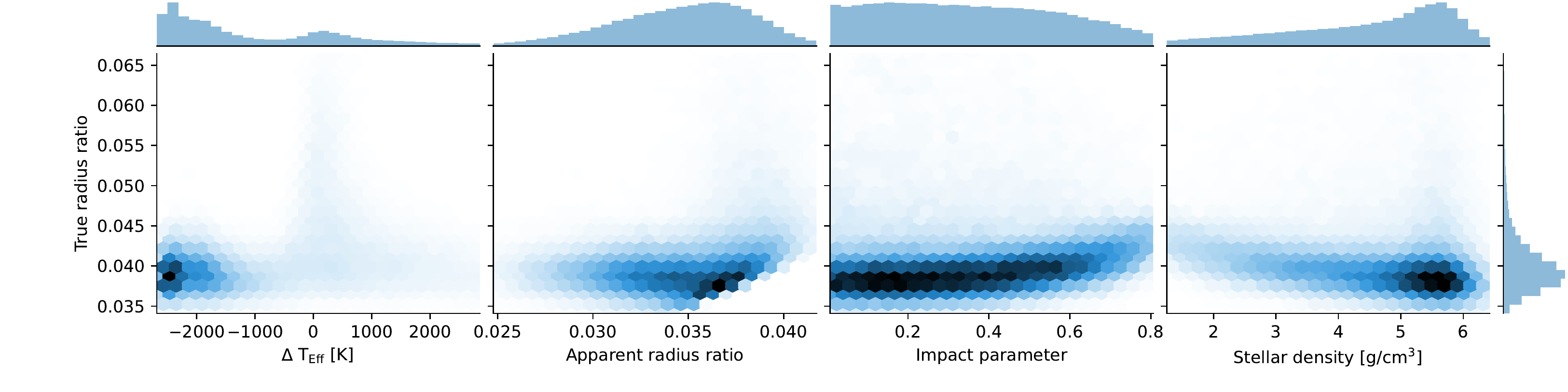}}
            \caption{\textit{Top four panels:}
            From left to right, \toicuatro posterior distributions of $k_{true}$ versus $\Delta T_{\rm eff}$, $k_{app}$, the impact parameter and the stellar density from the joint multicolor lightcurve analysis. \textit{Bottom four panels:} Same for \toidos.}
            
            \label{fig:Validation_plot_2081_4479}
    \end{center}
\end{figure*}

\section{Results and discussion}
\label{sec: Results}

In this section, we validated the planetary nature and discuss the properties of \toidosb and \toicuatrob. We evaluate the possible flux contamination in both systems by studying the posterior distributions of the true and apparent planet-to-star radius ratio ($k_{true}$, $k_{app}$) and the difference in effective temperature between the possible contaminant and the host star ($\Delta T_{\rm eff}$) parameters.

\begin{table}[h!]    
		\caption{Planetary parameters of \toidosb and \toicuatrob derived through the multicolor validation pipeline.}
		\centering
		\begin{tabular*}{\columnwidth}{@{\extracolsep{\fill}} lll}
			\hline\hline \\[1pt]
			\emph{} & \toicuatrob & \toidosb\\[2pt]
			\hline\\[0.01cm]         
			\centering
			$R_p$ [$\rm R_{\oplus}$] & $ 2.82^{+0.65}_{-0.63}$ & $ 2.04^{+0.50}_{-0.54}$\\[5pt]
			$k_{\rm app}$ & $ 0.0572^{+0.0024}_{-0.0017}$ & $ 0.0350^{+0.0032}_{-0.0041}$\\[5pt]
			$k_{\rm true}$ & $ 0.062^{+0.011}_{-0.004}$ & $ 0.0396^{+0.0039}_{-0.0024}$\\[5pt]
			$P_{\rm orb}$ [days] & $ 1.15890^{+0.00001}_{-0.00002}$ & $ 10.50534^{+0.00007}_{-0.00008}$\\[5pt]
			$a/R_{*}$ & $ 7.8^{+0.7}_{-1.4}$ & $ 30.3^{+1.9}_{-4.8}$\\[5pt]
			$a$ [AU] & $0.0164^{+0.0015}_{-0.0029}$ & $0.0752^{+0.0047}_{-0.0119}$\\[5pt]
			$b$ & $ 0.43^{+0.24}_{-0.30}$ & $ 0.35^{+0.27}_{-0.24}$\\[5pt]
			$i$ [deg] & $86.36_{-2.72}^{+2.49}$ & $89.34_{-0.73}^{+0.46}$\\[5pt]
			$T_{\rm c}$ [BJD] & $2459420.7578^{+0.0013}_{-0.0011}$ & $2458685.8996^{+0.0029}_{-0.0028}$\\[5pt]
			$T_{\rm eq}$ [$\rm K$] & $861^{+64}_{-103}$ & $488^{+28}_{-52}$\\[5pt]
			$M_{p}$ [$M_{\oplus}$]$^{1}$ & $8.3_{-4.1}^{+8.0}$ & $5.0_{-2.4}^{+4.8}$\\[5pt]
			$F$ [$F_{\oplus}$] & $92.5^{+17.5}_{-33.3}$ & $8.1^{+1.1}_{-2.6}$\\[5pt]
			$K$ [$\rm m/s$]$^{2,3}$ & 7.12 & 1.72\\[5pt]
			\hline
		\end{tabular*}
		\tablefoot{$^{1}$Predicted masses using \texttt{Forecaster} \citep{ChenKipping17_Forecaster} empirical mass-radius relation. $^{2}$Taken from the TESS Input Catalog \citep[\texttt{TICv8,}][]{Stassun2019_TIC} $^{3}$Predicted RV semiamplitude.}
		\label{tab:planetary_parameters}  
	\end{table}

\subsection{\toicuatro}

We show the phase-folded combined MuSCAT2, TESS and SINISTRO photometric datasets of \toicuatrob with the best lightcurve models in Figure~\ref{fig:M2_LC}, Figure~\ref{fig:TESS_LCs} and Figure~\ref{fig:LCO_LC}, respectively. Also, we show in Table~\ref{tab:planetary_parameters} the derived stellar and planetary parameters from our multicolor validation pipeline. The corner-plot showing the parameter posterior distributions can be found in Figure~\ref{fig:Corners_TOI4479}.

\begin{figure*}[h!]
    
    \begin{center}
        \centerline{\includegraphics[width=0.5\textwidth]{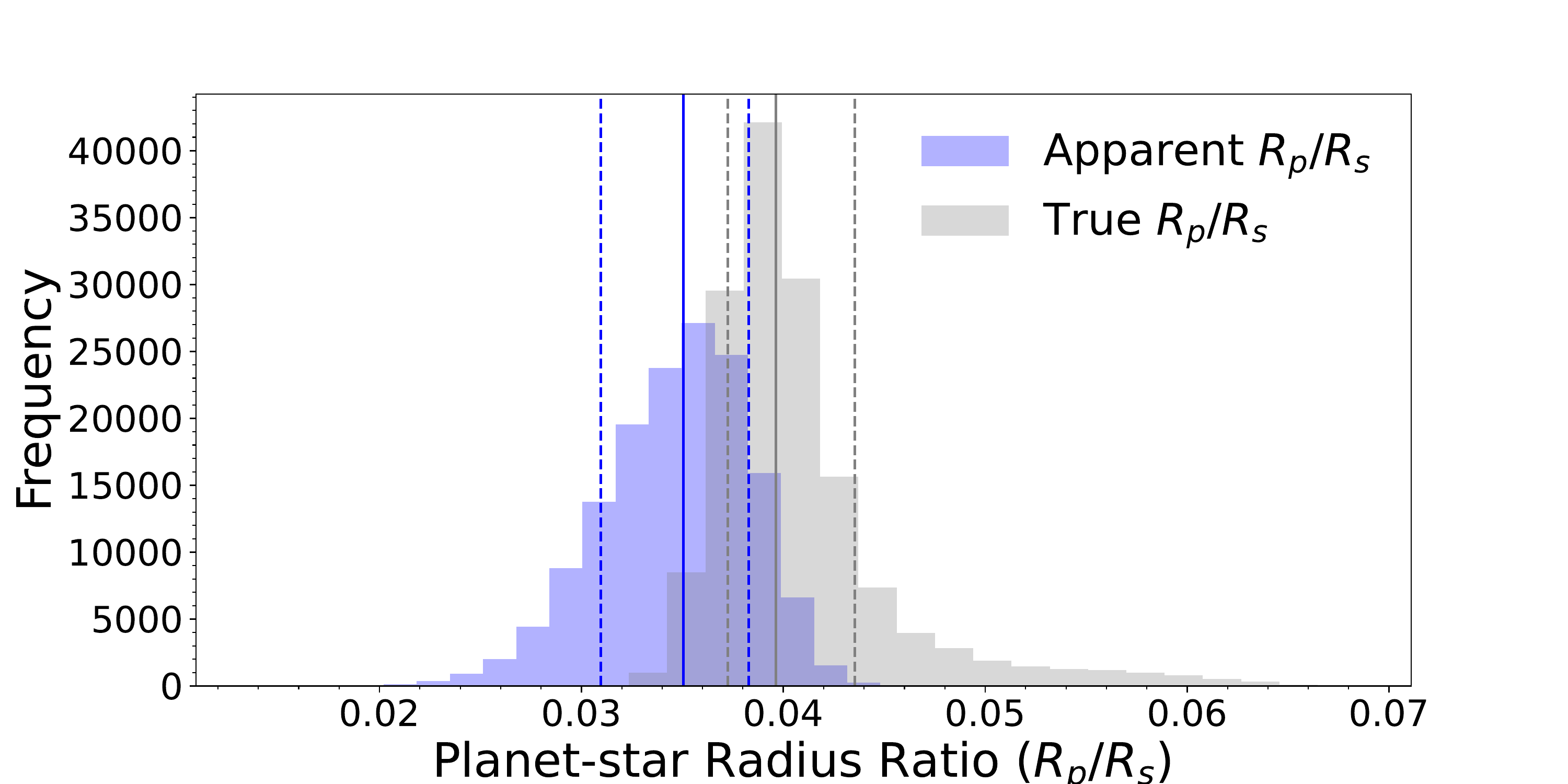}
        \includegraphics[width=0.5\textwidth]{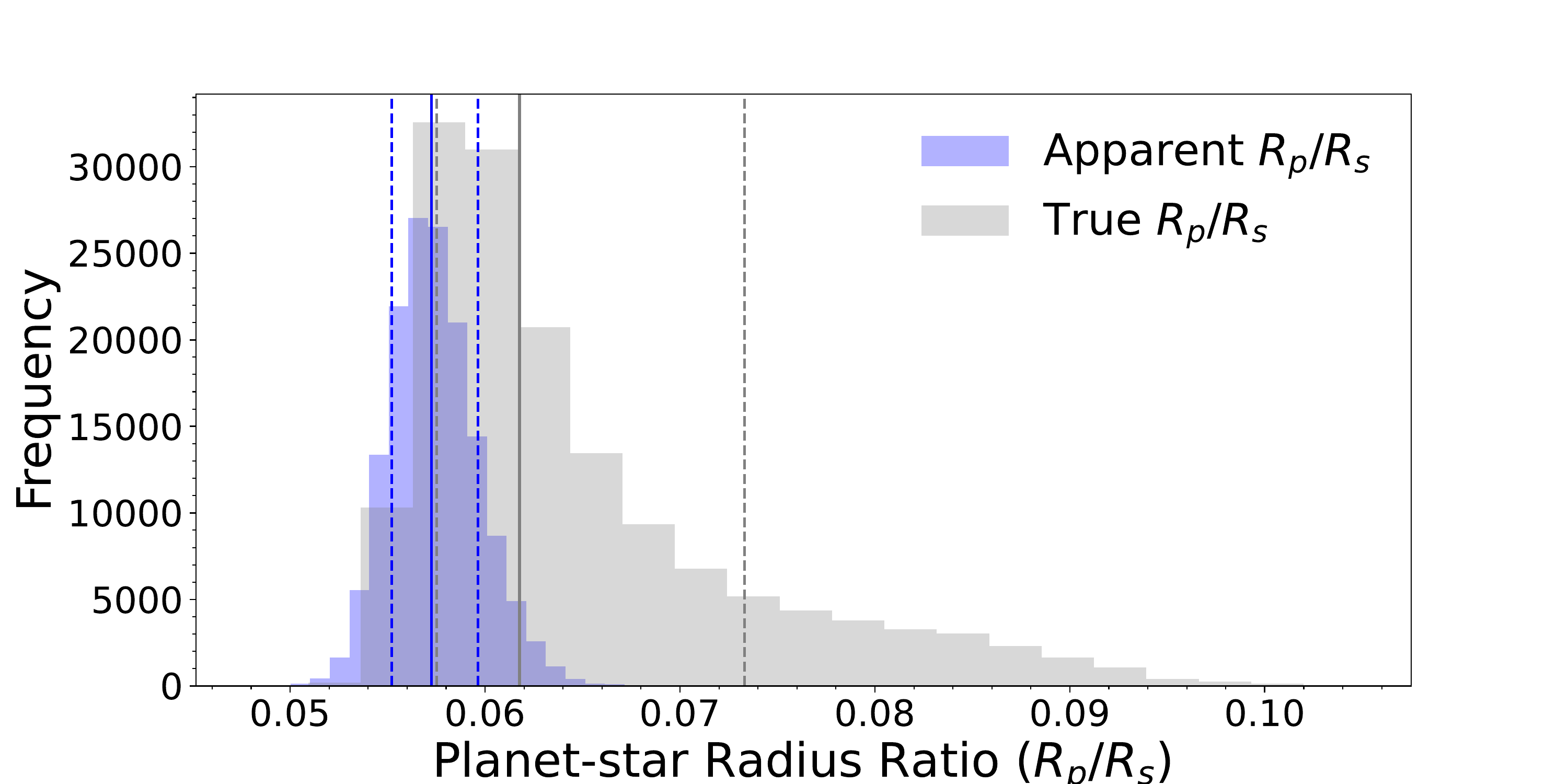}}
        \centerline{\includegraphics[width=0.5\textwidth]{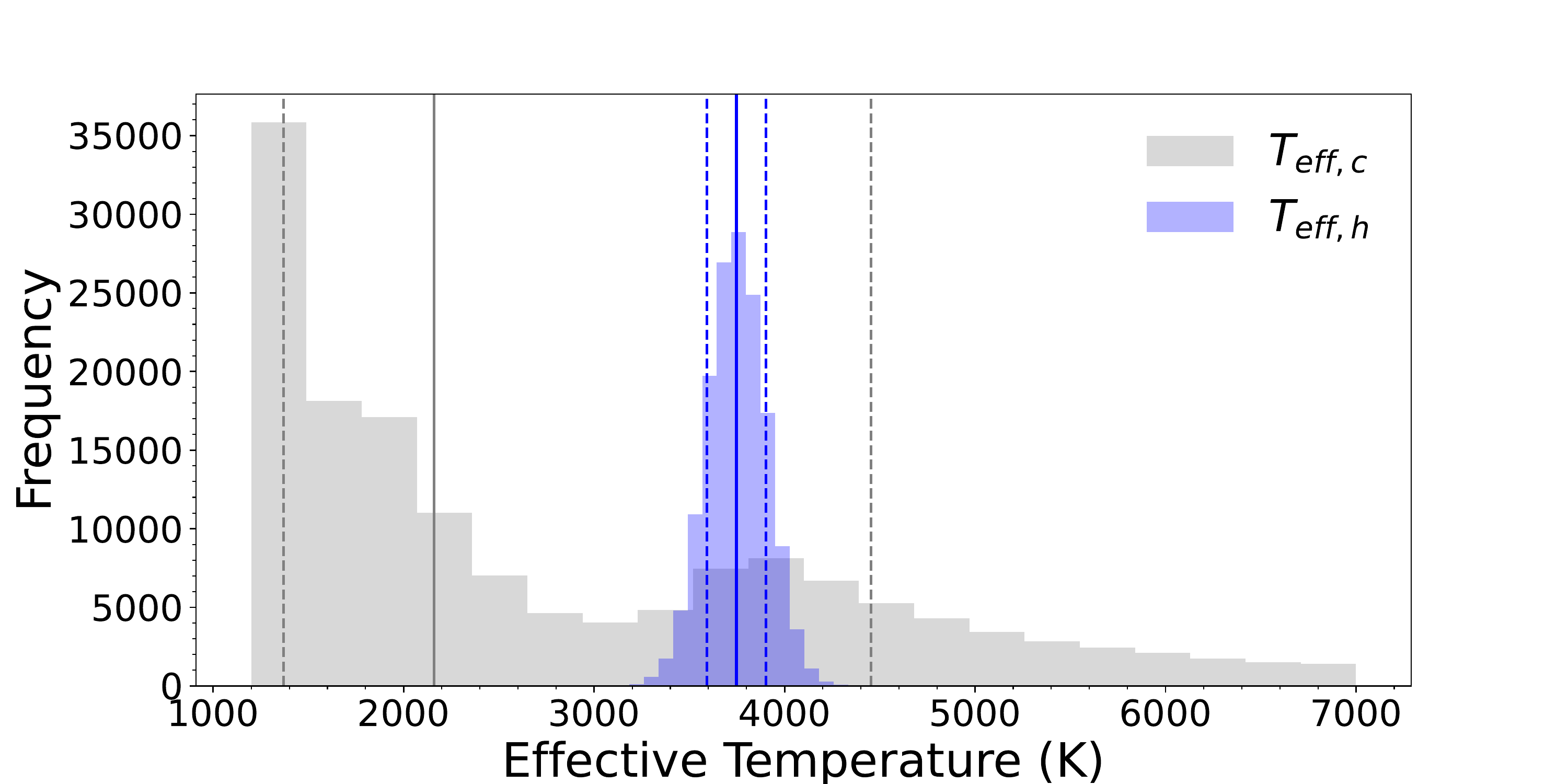}
        \includegraphics[width=0.5\textwidth]{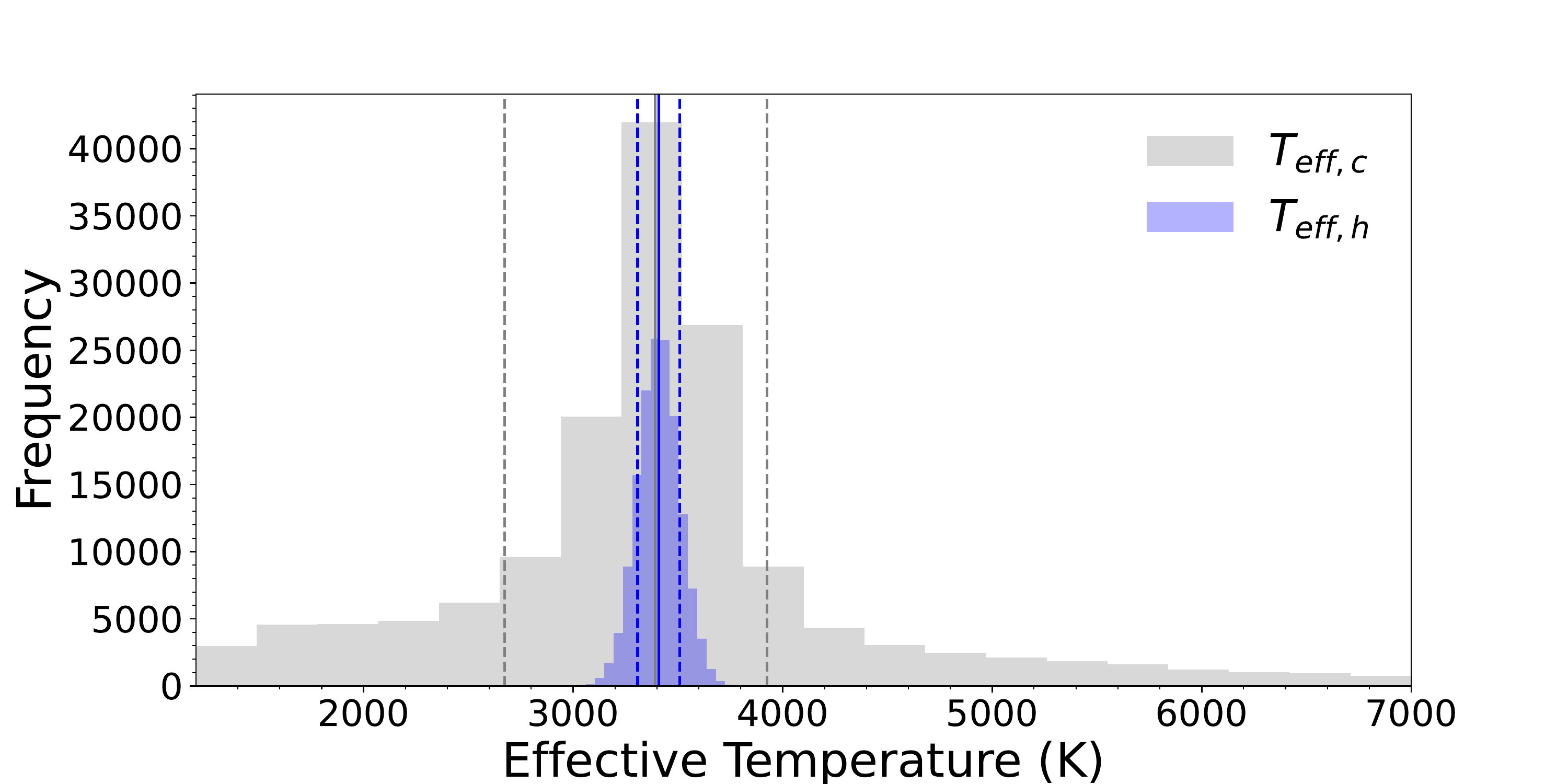}}
        
            \caption{\textit{Upper panels:} Posterior distributions of the true radius-ratio ($k_{\rm true}$) and the apparent radius-ratio ($k_{\rm app}$) for \toidos system (left) and \toicuatro system (right). \textit{Lower panels:} Posterior distributions of the effective temperature of the host star ($T_{\rm eff,~h}$) and the effective temperature of the contaminant ($T_{\rm eff,~c}$) for \toidos system (left) and \toicuatro system (right). The solid lines show the median of each distribution and the dashed lines show the lower and upper $1\sigma$.}
            \label{fig:Distributions_plot_2081_4479}
    \end{center}
\end{figure*}

To evaluate the possible flux contamination, in Figure~\ref{fig:Validation_plot_2081_4479} we show the posterior distribution of the true radius-ratio ($k_{true}$) as a function of the difference in effective temperature between the contaminant and the host star ($\Delta T_{\rm eff}$), the apparent radius-ratio ($k_{app}$), the impact parameter and the stellar density. We also show a comparison among the posterior distributions of the apparent and true radius ratio as well as the effective temperatures of the host star and contaminant for \toicuatro system in Figure~\ref{fig:Distributions_plot_2081_4479}.

For \toicuatro, we found $k_{true}$ to be close in value to $k_{app}$ (Figure~\ref{fig:Validation_plot_2081_4479}), implying a very low degree of flux contamination from a possible companion. Thus, considering the contamination negligible, we derived the size of the companion from the $k_{app}$, leading to a $2.82 ^{+0.65}_{-0.63} \rm R_{\oplus}$ sized object. Moreover, the posterior distribution of $\Delta T_{\rm eff}$ is centered around 0, meaning that the effective temperature of the possible contaminant would be the same as that of the host star, and the posterior distribution of the impact parameter implies a non-grazing transit. Thus, we can validate \toicuatrob as a sub-Neptune sized planet orbiting around an M dwarf with a period of $1.15890^{+0.00001}_{-0.00002}$ days.

\begin{figure*}[h!]
    
    \begin{center}
        
        \centerline{\includegraphics[width=\textwidth]{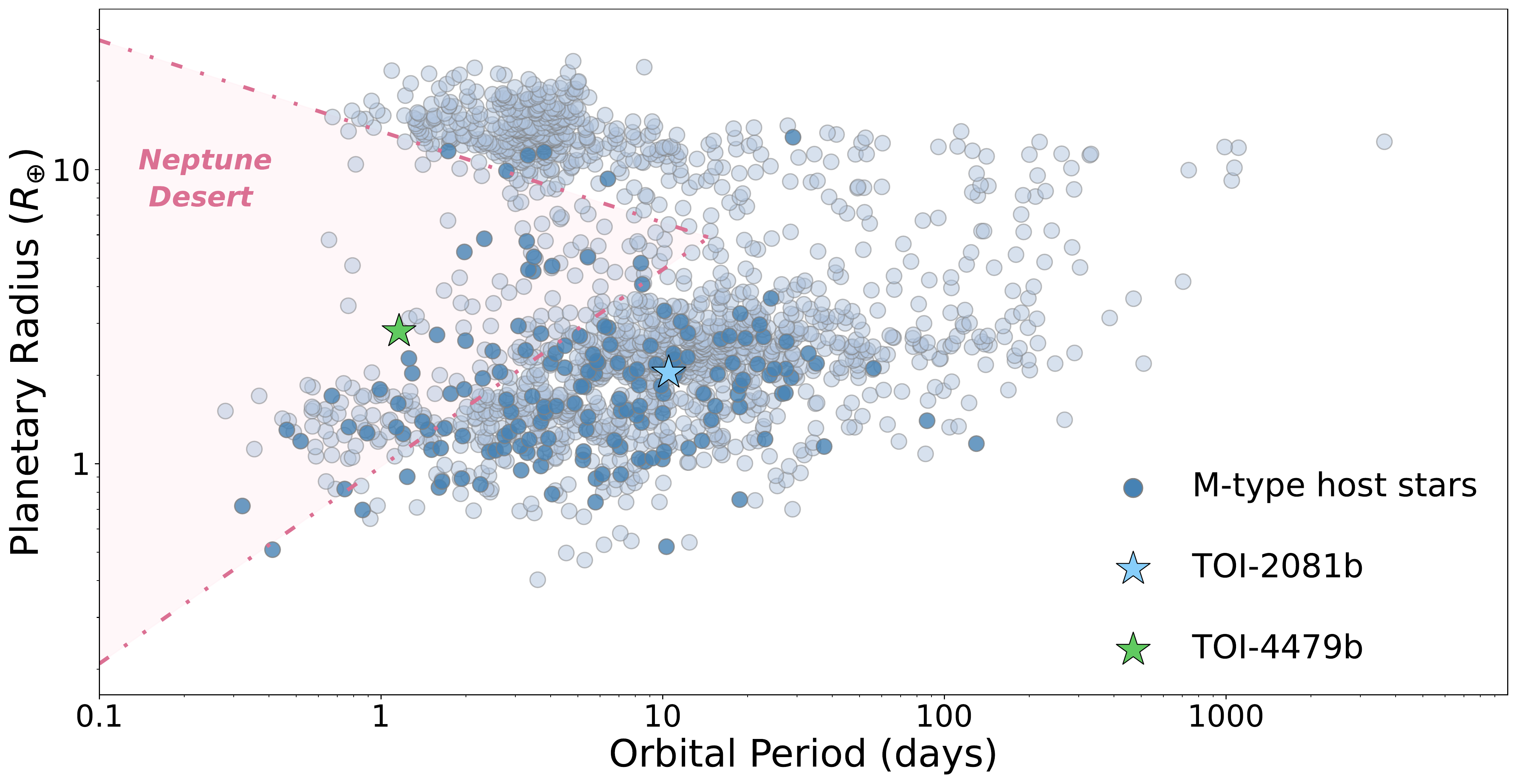}}
            \caption{\toicuatrob and \toidosb in the Period-Radius diagram compared to all the confirmed planets to date with a radius uncertainty below 10\%. The dark blue dots show the planets around M-type stars and the dashed-dotted lines show the Neptune desert boundaries in the Period-Radius plane derived by \cite{Mazeh2016}.}
            \label{fig:Demographics}
    \end{center}
\end{figure*}

In Figure~\ref{fig:Demographics} we compare \toicuatrob with the sample of confirmed planets around M dwarfs known to date in the Period-Radius plane. We also compare \toicuatrob with the entire population of confirmed planets with a radius uncertainty below $10\%$, showing that \toicuatrob lays in an underpopulated region of the Period-Radius plane known as the Neptune desert. This deserted region contrasts with the highly populated regions of hot-Jupiters ($R_{p}>10~R_{\oplus}$) and ultra-short period (USP) rocky planets ($R_{p}<2R_{\oplus}$) located above and below the Neptune desert, respectively. We have plotted with dashed-dotted lines in Figure~\ref{fig:Demographics} the lower and upper boundaries among these regions as derived by \cite{Mazeh2016}. The dearth of short-period Neptune-sized planets has been widely studied in the literature and several formation mechanisms (e.g., photo-evaporation, high-eccentricity migration, in-situ formation) have been considered to explain the causes of the Neptune desert region \citep{Sanchis-Ojeda14,Mazeh2016,Lundkvist16,Lopez17,OwenWu17,OwenLai18}. 

Although \toicuatrob has a slightly longer period that the common definition of USP planets ($P_{\rm orb}$<1~day), given their shared properties we associate it here to this population. \toicuatrob joins a small group of known USP intermediate-sized planets inhabiting the Neptune desert, e.g. LTT 9779b \citep{Jenkins20}, LP 714-47 b \citep{Dreizler20}, HATS-37Ab \citep{Jordan20}, HATS-38b \citep{Jordan20}, TOI-824 b \citep{Burt20}, TOI-849b \citep{Armstrong20}, TOI-132 b \citep{Diaz20}, TOI-674b \citep{Murgas2021}. The existence of such uncommon planets has been interpreted as a consequence of the photo-evaporation produced in short time-scales by the strong stellar irradiation experienced by some low-mass planets, which are unable to retain the H/He envelope \citep{Jenkins20}. However, \toicuatrob is among the biggest USP planets in the desert, meaning that it is still retaining an appreciable fraction of its volatile envelope in an early stage of the stripping process. We find \toicuatrob to be the biggest USP planet orbiting around an M dwarf known to date. We used \texttt{Forecaster}\footnote{\url{https://github.com/chenjj2/forecaster}} \citep{ChenKipping17_Forecaster} to predict the plausible mass of \toicuatrob, which is 
$M_p = 8.3_{-4.1}^{+8.0} \, M_{\oplus}$. Note that the error bars on the 
mass are dominated by the intrinsic spread of the mass-radius 
distribution for Neptunian planets relative to the simple power law 
relation.
We evaluated the prospects to spectroscopically investigate the 
atmosphere of \toicuatrob by computing the transmission spectroscopy 
metric (TSM), as defined by \cite{kempton2018}. The TSM is inversely 
proportional to the planetary mass, and using the forecasted mass range, the 
TSM ranges between 26 and 198 with a peak at $\mathrm{TSM} \sim 75$. 
According to \cite{kempton2018}, Neptune-sized planet with 
$\mathrm{TSM} \gtrsim 90$ are the highest-priority candidates for 
transmission spectroscopy with the JWST, where the ranking is based on 
the predicted S/N of atmospheric detections. However, individual planets with $\mathrm{TSM}
< 90$ can still be extremely suitable candidates for transmission spectroscopy with 
the JWST, based on scientific merit. This applies to  \toicuatrob due to the fact that it orbits around an M star and its located in the Neptune desert.

\subsection{\toidos}

We show the phase-folded combined MuSCAT2 and TESS photometric datasets of \toidosb together with the best lightcurve models in Figure~\ref{fig:M3_LC} and Figure~\ref{fig:TESS_LCs}, respectively. Also, we show in Table~\ref{tab:planetary_parameters} the derived stellar and planetary parameters from our multicolor validation pipeline. The corner-plot showing the parameter posterior distributions can be found in Figure~\ref{fig:Corners_TOI2081}.

To evaluate the possible flux contamination, in Figure~\ref{fig:Validation_plot_2081_4479} we show the posterior distribution $k_{true}$ as a function of the difference in $\Delta T_{\rm eff}$, $k_{app}$, the impact parameter and the stellar density. The comparison among the posterior distributions of the apparent and true radius ratio as well as the effective temperatures of the host star and contaminant for \toidos system is shown in Figure~\ref{fig:Distributions_plot_2081_4479}.

For \toidos, we found $k_{true}$ to be close in value to $k_{app}$ (Figure~\ref{fig:Validation_plot_2081_4479}), implying a very low degree of flux contamination from the companion. Considering the contamination negligible, we derived the size of the companion from the $k_{app}$, leading to a $2.04 ^{+0.49}_{-0.54}~\rm R_{\oplus}$ sized object. Thus, we can validate \toidosb as a super-Earth sized planet orbiting around an M dwarf with a period of $ 10.50534^{+0.00007}_{-0.00008}$ days.

\toidosb is also included in the radius-period diagram in Figure~\ref{fig:Demographics}. We find \toidosb to be a temperate super-Earth, in a well-populated parameter space region both around M dwarfs and earlier stellar type hosts.

\toidosb ($a\sim0.07~\rm AU$) is orbiting within the inner edge of the habitable zone of its star (we get a conservative habitable zone of $[0.16 \pm 0.02,0.34 \pm 0.04]$ AU.), and in a tidally locked regime. Assuming, a zero Albedo and a cloud free atmosphere without greenhouse gases, the temperature of the day side is estimated around $T_\mathrm{day} \approx 680 K$. Using \texttt{Forecaster}, we estimated a mass of $M_p = 5.0_{-2.4}^{+4.8} \, M_{\oplus}$. The corresponding TSM ranges between 14 and 89 with peak at $\mathrm{TSM} \sim 35$.

\section{Conclusions}
\label{sec: Conclusions}

By using multi-color photometric observations with MuSCAT2, MuSCAT3 and LCOGT~1m we determined that the degree of contamination by a possible nearby contaminant is negligible in both the \toicuatrob and the \toidosb systems, and validated their planetary nature. \toicuatrob is a sub-Neptune sized planet ($R_{p}=2.82 ^{+0.65}_{-0.63}~\rm R_{\oplus}$) and \toidosb is a super-Earth sized planet ($R_{p}=2.04 ^{+0.49}_{-0.54}~\rm R_{\oplus}$). Both planets orbit around M dwarf host stars with orbital periods of $10.50534\pm0.00007$~days and $1.15890^{+0.00002}_{-0.00001}$~days, respectively.

We also found that the \toicuatrob lays in the Neptune desert and joins a small sample of ($\sim8$) short-period intermediate-sized planets, with \toicuatrob being the biggest USP planet orbiting around an M dwarf known to date. Thus, this planet is an interesting target for future radial velocity observations (which will require very large telescope apertures) and atmospheric studies, as its full characterization may provide significant observational constraints for planet formation and evolution theories.

\begin{acknowledgements}
      
      We thank the anonymous referee for insightful suggestions, which added the clarity of this paper.
      
      Funding for the TESS mission is provided by NASA's Science Mission Directorate.
      
      We acknowledge the use of public TESS data from pipelines at the TESS Science Office and at the TESS Science Processing Operations Center. Resources supporting this work were provided by the NASA High-End Computing (HEC) Program through the NASA Advanced Supercomputing (NAS) Division at Ames Research Center for the production of the SPOC data products.
      
      We acknowledge the use of public TESS data from pipelines at the TESS Science Office and at the TESS Science Processing Operations Center.
      
      This research has made use of the Exoplanet Follow-up Observation Program website, which is operated by the California Institute of Technology, under contract with the National Aeronautics and Space Administration under the Exoplanet Exploration Program.
      
      This paper includes data collected by the TESS mission that are publicly available from the Mikulski Archive for Space Telescopes (MAST).
      
      This paper is based on observations made with the MuSCAT2 instrument, developed by ABC, at Telescopio Carlos Sánchez operated on the island of Tenerife by the IAC in the Spanish Observatorio del Teide. 
      
      This paper is based on observations made with the MuSCAT3 instrument, developed by the Astrobiology Center and under financial supports by JSPS KAKENHI (JP18H05439) and JST PRESTO (JPMJPR1775), at Faulkes Telescope North on Maui, HI, operated by the Las Cumbres Observatory.
      
      This work makes use of observations from the LCOGT network. Part of the LCOGT telescope time was granted by NOIRLab through the Mid-Scale Innovations Program (MSIP). MSIP is funded by NSF.
      
      Based on observations made with the Nordic Optical Telescope, owned in collaboration by the University of Turku and Aarhus University, and operated jointly by Aarhus University, the University of Turku and the University of Oslo, representing Denmark, Finland and Norway, the University of Iceland and Stockholm University at the Observatorio del Roque de los Muchachos, La Palma, Spain, of the Instituto de Astrofisica de Canarias.
      
      The data presented here were obtained in part with ALFOSC, which is provided by the Instituto de Astrofisica de Andalucia (IAA) under a joint agreement with the University of Copenhagen and NOT.
      
      This work made use of \texttt{tpfplotter} by J. Lillo-Box (publicly available in www.github.com/jlillo/tpfplotter), which also made use of the python packages \texttt{astropy}, \texttt{lightkurve}, \texttt{matplotlib} and \texttt{numpy}.
      
      E. E-B. acknowledges financial support from the European Union and the State Agency of Investigation of the Spanish Ministry of Science and Innovation (MICINN) under the grant PRE2020-093107 of the Pre-Doc Program for the Training of Doctors (FPI-SO) through FSE funds. 
      
      G. M. has received funding from the European Union's Horizon 2020 research and innovation programme under the Marie Sk\l{}odowska-Curie grant agreement No. 895525.
      
      C.D. D. acknowledges support provided by the NASA Exoplanets Research Program (XRP) via grant 80NSSC20K0250.
      
      J.K. gratefully acknowledge the support of the Swedish National Space Agency (SNSA; DNR 2020-00104).
      
      R.L. acknowledges financial support from the Spanish Ministerio de Ciencia e Innovación, through project PID2019-109522GB-C52, and the Centre of Excellence "Severo Ochoa" award to the Instituto de Astrofísica de Andalucía (SEV-2017-0709).
      
      A.A.B. and M.V.G. acknowledge the support of Ministry of Science and Higher Education of the Russian Federation under the grant 075-15-2020-780(N13.1902.21.0039).
      
      M.T. is supported by JSPS KAKENHI grant Nos.18H05442, 15H02063, and 22000005.
      
      M.R.Z.O. acknowledges financial support from the Spanish Ministerio de Ciencia e Innovación through project PID2019-109522GB-C51.
      
      This work is partly supported by JSPS KAKENHI Grant Numbers 22000005, JP15H02063, JP17H04574, JP18H05439, P18H05442, JP20K14518, JP20K14521, JP21K13975, JP21K20376, Grant-in-Aid for JSPS Fellows Grant Number JP20J21872, JST CREST Grant Number JPMJCR1761, and the Astrobiology Center of National Institutes of Natural Sciences (NINS) (Grant Numbers AB022006, AB031010, AB031014). 
\end{acknowledgements}

\bibliographystyle{aa} 
\bibliography{Biblio} 

\begin{appendix}

\onecolumn{
\section{Additional plots}
    \begin{figure}[h!]
    \centering
    
    \includegraphics[width=\textwidth]{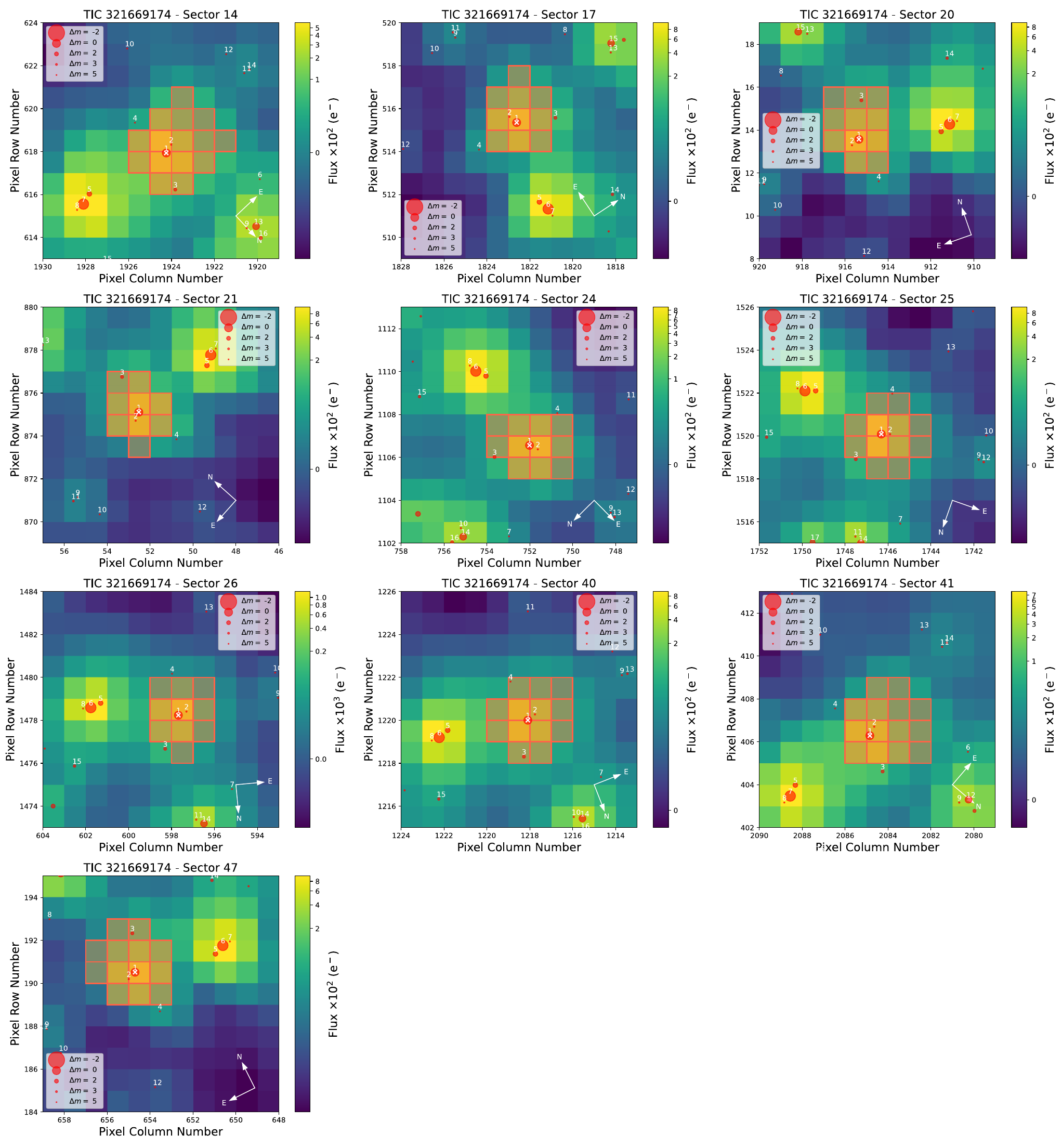}
    \caption{TESS target pixel file images of \toidos observed in Sectors 14, 17, 20, 21, 24, 25, 26, 40, 41 and 47. The red circles show the sources in the field identified by the \emph{Gaia DR2} catalogue with scaled magnitudes. The position of the targets is indicated by white crosses and the mosaic of orange squares show the mask used by the pipeline to extract photometry. These plots were made using \texttt{tpfplotter} \citep{Aller2020_tpfplotter}.}
    \label{fig:TPF_2081_mosaic}
\end{figure}

    \begin{figure}[h!]
    \centering
    
    \includegraphics[width=\textwidth]{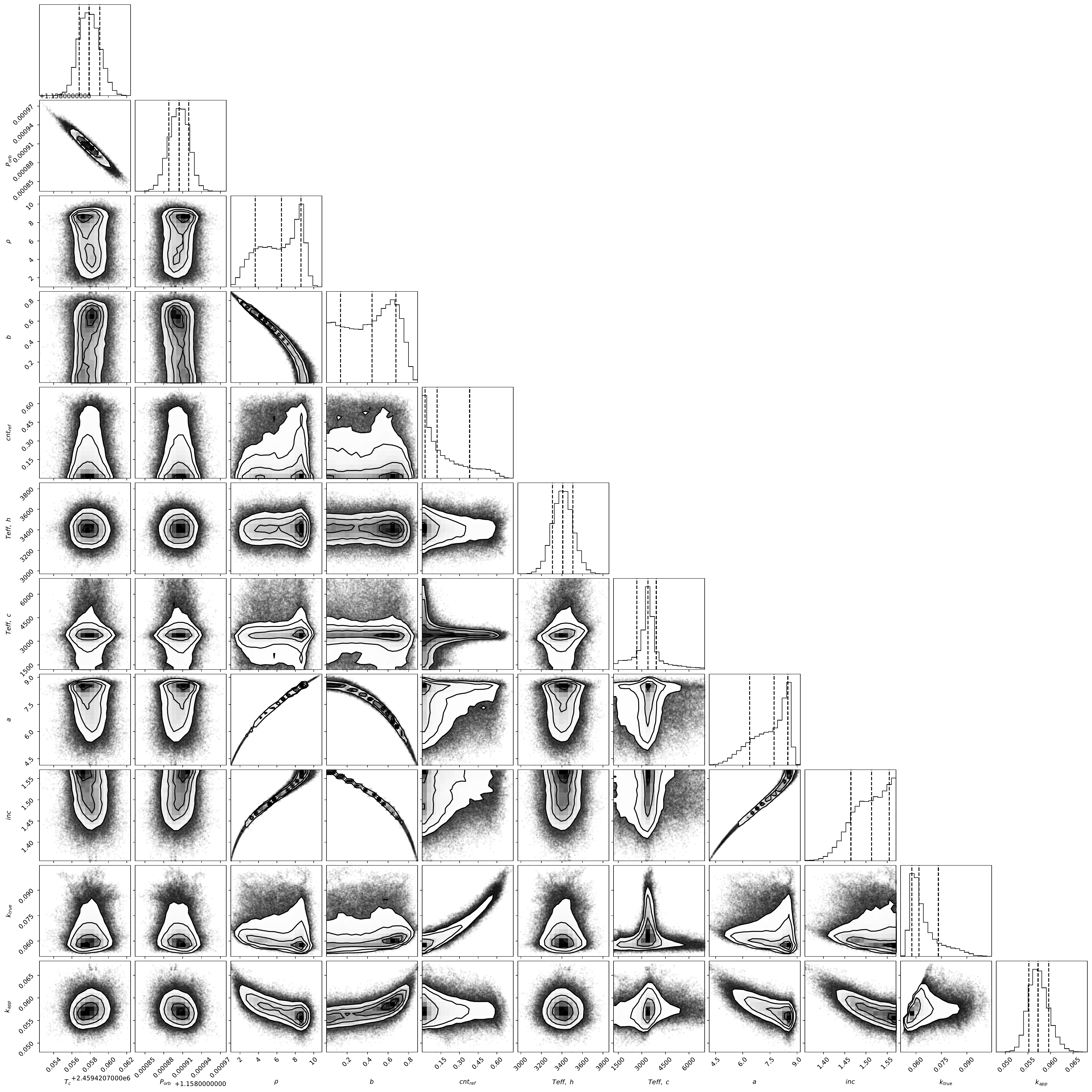}
    \caption{Corner plot of the posterior distributions obtained through the multicolor validation pipeline for \toicuatro.}
    \label{fig:Corners_TOI4479}
\end{figure}

    \begin{figure}[h!]
    \centering
    
    \includegraphics[width=\textwidth]{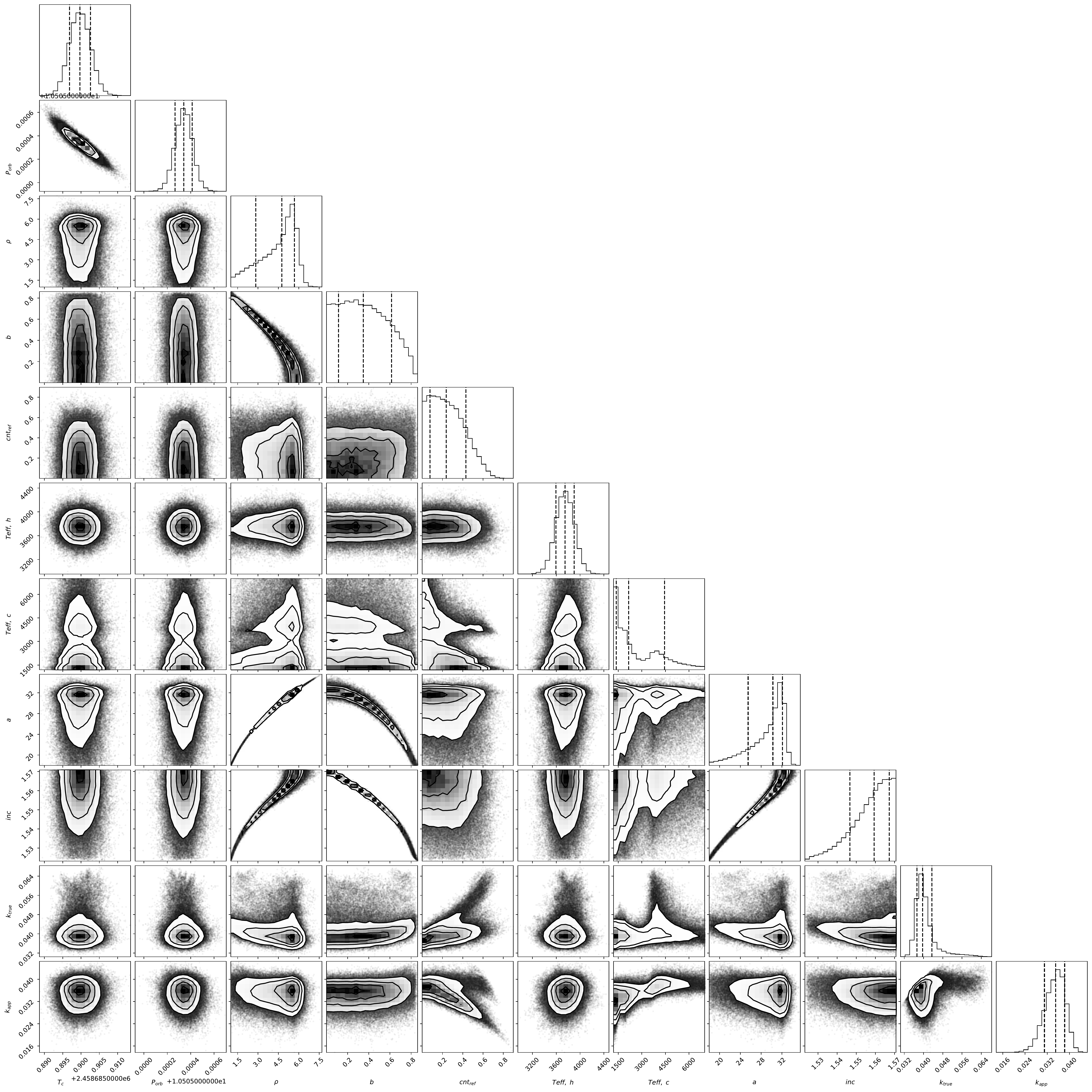}
    \caption{Corner plot of the posterior distributions obtained through the multicolor validation pipeline for \toidos.}
    \label{fig:Corners_TOI2081}
\end{figure}}

\end{appendix}

\end{document}